\documentclass[a4paper]{report}
\usepackage[utf8]{inputenc}
\usepackage[T1]{fontenc}
\usepackage{RJournal}
\usepackage{amsmath,amssymb,array}
\usepackage{booktabs}

\usepackage{graphicx, subfigure} 
\usepackage{multirow}            
\usepackage{hyperref}

\usepackage{amsfonts} 
\usepackage{algorithm}
\usepackage{algorithmic}
\usepackage[capitalize]{cleveref} 

\usepackage{color,soul}
\usepackage{tikz}
\newcommand{\tikzcircle}[2][red,fill=red]{\tikz[baseline=-0.5ex]\draw[#1,radius=#2] (0,0) circle ;}
\definecolor{chestnut}{rgb}{0.8, 0.36, 0.36} 
\definecolor{apricot}{rgb}{0.98, 0.81, 0.69} 
\definecolor{burlywood}{rgb}{0.87, 0.72, 0.53}
\definecolor{iceberg}{rgb}{0.44, 0.65, 0.82} 
\definecolor{applegreen}{rgb}{0.55, 0.71, 0.0} 
\definecolor{bronze}{rgb}{0.8, 0.5, 0.2} 
\definecolor{beaublue}{rgb}{0.74, 0.83, 0.9} 
\definecolor{celadon}{rgb}{0.67, 0.88, 0.69} 
\definecolor{carrotorange}{rgb}{0.93, 0.57, 0.13} 
\definecolor{amber}{rgb}{1.0, 0.75, 0.0} 
\definecolor{bubblegum}{rgb}{0.99, 0.76, 0.8} 

\begin{document}

\sectionhead{Contributed research article}
\volume{XX}
\volnumber{YY}
\year{20ZZ}
\month{AAAA}

\begin{article}
\title{\pkg{netgwas}: An R Package for Network-Based Genome Wide Association Studies}
\author{by Pariya Behrouzi, Danny Arends and Ernst C. Wit}

\maketitle

\abstract{  
Graphical models are a powerful tool in modelling and analysing complex biological associations in high-dimensional data. The R-package \pkg{netgwas} implements the recent methodological development on copula graphical models to (i) construct linkage maps, (ii) infer linkage disequilibrium networks from genotype data, and (iii) detect high-dimensional genotype-phenotype networks. The \pkg{netgwas} learns the structure of networks from ordinal data and mixed ordinal-and-continuous data. 
Here, we apply the functionality in \pkg{netgwas} to various multivariate example datasets taken from the literature to demonstrate the kind of insight that can be obtained from the package. We show that our package offers a more realistic association analysis than the classical approaches, as it discriminates between direct and induced correlations by adjusting for the effect of all other variables while performing pairwise associations. This feature controls for spurious interactions between variables that can arise from conventional approaches in a biological sense. The \pkg{netgwas} package uses a parallelization strategy on multi-core processors to speed-up computations.

}

\section{Introduction}
 \label{chap4:into}
\begin{figure}[t]
	\hspace{-0.5cm}
	\includegraphics[width=1.08\textwidth]{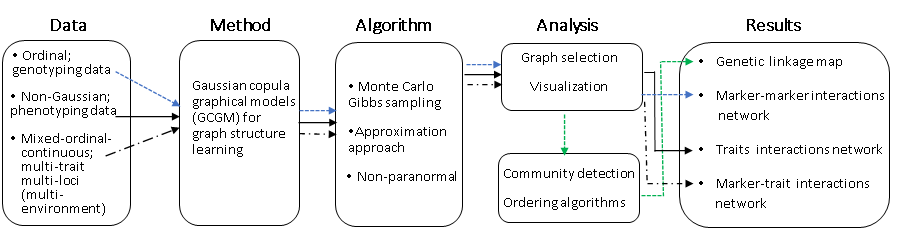} 
	\caption{Configuration of the \pkg{netgwas} package. Depending on the data type different functions can be used. Each color represents one main function of the package. The three main functions are: i) {\tt netmap()} (in green) constructs linkage maps for bi-parental species with any ploidy level, ii) {\tt netsnp()} (in blue) detects the conditional dependent short- and long-range linkage disequilibrium structure of genomes and iii) {\tt netphenogeno()} (in black) reconstructs  association networks between phenotypes and genetic markers.}
	\label{Flowchart}
\end{figure}

Graphical models are commonly used in statistics and machine learning to model complex dependency structures in multivariate data \citep{lauritzen1996graphical, hartemink2000using, lauritzen2003graphical, jordan2004graphical, friedman2004inferring, dobra2004sparse, edwards2010selecting, behrouzi2018dynamic, vinciotti2022bayesian} where each node in the graph represents a random variable and edges represent conditional dependence relationships between pairs of variables. Therefore, the absence of an edge between two nodes indicates that the two variables are conditionally independent. The \pkg{netgwas} package contains an implementation of undirected graphical models to address the three key and interrelated goals in genetics association studies: (i) building linkage maps, (ii) reconstructing linkage disequilibrium networks, and (iii) detecting genotype-phenotype networks (see Fig \ref{Flowchart}). Below we provide a brief introduction for each section of \pkg{netgwas}. 

A linkage map describes the linear order of genetic markers within linkage groups (chromosomes). It is the first requirement for estimating the genetic background of phenotypic traits in quantitative trait loci (QTL) studies and are commonly used in QTL studies to link phenotypic traits to the underlying genetics of the population. In practice, many software packages for performing QTL analysis require linkage maps \citep{lander1987mapmaker, broman2003r, yang2008qtlnetwork, taylor2011r, huang2012dlmap, broman2019r}. 
Most organisms are categorized as diploid or polyploid by comparing the copy number of each chromosome. Diploids have two copies of each chromosome (like humans). Polyploid organisms have more than two copies of each chromosome (like most crops). Polyploidy is common in plants and in different crops such as apple, potato, and wheat, which contain three (triploid), four (tetraploid), and six (hexaploid) copies from each of their chromosomes, respectively. Despite the importance of polyploids, statistical tools for their map construction are underdeveloped \citep{grandke2017pergola, bourke2018polymapr}. Most software packages such as {\tt R/qtl} \citep{broman2003r}, {\tt OneMap} \citep{margarido2007onemap}, {\tt Pheno2Geno} \citep{zych2015pheno2geno}, and {\tt MST{\scriptsize MAP}} \citep{wu2008efficient, taylor2017r} contain functionality to only construct linkage maps for diploid species. Packages such as {\tt MAPMAKER} \citep{lander1987mapmaker}, {\tt TetraploidSNPMap} \citep{hackett2017tetraploidsnpmap}, and {\tt polymapR} \citep{bourke2018polymapr} have the functionality to construct polyploid linkage maps but focus mainly on a specific type of polyploid species (e.g. tetraploids). All the aforementioned packages contain methods that use pairwise estimation of recombination frequencies and LOD (logarithm of the odds ratio) scores \citep{wang2016potential} that often require manual interaction. This often leads to an underpowered approach and confounding of correlated genotypes by failing to correct for intermediate markers \citep{behrouzi2019novo}. In contrast, the \pkg{netgwas} R package uses a multivariate approach to construct linkage maps for diploid and polyploid species in a unified way. This is achieved by utilising the pairwise Markov property between any two genetic markers and constructing the linkage map by simultaneously assessing the complete set of pairwise comparisons. This often leads to an improved marker order over more conventional methods.

The linkage map,  which can be constructed using the first key function of \pkg{netgwas} in Fig \ref{Flowchart}, provides the genetic basis for the second key function which detects the patterns of linkage disequilibrium and segregation distortion in a population. Segregation distortion (SD) refers to any deviation from expected segregation ratios based on Mendelian rules of inheritance. 
And linkage disequilibrium (LD) refers to non-random relations between loci (locations) on the same or different chromosomes.
Revealing the structures of LD is important for association mapping study as well as for studying the genomic architecture of a genome. Various methods have been published in the literature for measuring statistical associations between alleles at different loci, for instance see \citet{hedrick1987gametic, mangin2012novel, clarke2011basic, bush2012chapter, kaler2020comparing}. Most of these measures are based on an exhaustive genome scan for pairs of loci and the $r^2$ measure,  the square of the loci correlation. The drawback of such approaches is that association testing in the genome--scale is underpowered, so that weak long--range LD will go undetected. Furthermore, they do not simultaneously take the information of more than two loci into account to make full and efficient use of modern multi--loci data.
The \pkg{netgwas} package contains functionality to estimate pairwise interactions between different loci in a genome while adjusting for the effect of other loci to efficiently detect short- and long--range LD patterns in diploid and polyploid species \citep{behrouzi2019detecting}. Technically, this requires estimating a sparse adjacency matrix from multi-loci genotype data, which usually contains a large number of markers (loci), where the number of markers can far exceed the number of individuals. The non-zero patterns of the adjacency matrix created by the functions in \pkg{netgwas} shows the structures of short-- and long--range LD of the genome. The strength of associations between distant loci can be calculated using partial correlations. Furthermore, the methods implemented in \pkg{netgwas} already account for the correlation between markers, while associating them to each other and thereby avoids the problem of population structure (that is physically unlinked markers are correlated). 

A major problem in genetics is the association between genetic markers and the status of a disease (trait or phenotype). Genome-wide association studies are among the most important approaches for further understanding of genetics underlying complex traits \citep{welter2013nhgri, kruijer2020reconstruction}. However, in genome-wide association methods genetic markers are often tested individually for association with the phenotype. Since genome-wide scans analyze thousands or even millions of markers, the issue of multiple testing is usually addressed by using a stringent significance threshold \citep{panagiotou2011should}. Such methods work only if the associations are strong enough to pass the stringent threshold. However, even if that is the case, this type of analysis has several limitations, which have been discussed extensively in the literature \citep{ hoggart2008simultaneous, he2010variable, rakitsch2012lasso, buzdugan2016assessing}. 
Particularly, the main issue of this type of analysis is when we test the association of the phenotype to each genetic marker individually, and ignore the effects of all other genetic markers. This leads to failures in the identification of causal loci. 
If we consider two correlated loci, of which only one is causal for the phenotype, both may show a marginal association, but only the causal locus will be detected by our method. The methods implemented in the \pkg{netgwas} package tackle this issue by using Gaussian copula graphical model \citep{klaassen1997efficient, hoff2007extending}, which accounts for the correlation between markers, while associating them to the phenotypes. As it is shown in \citet{klasen2016multi}, this key feature avoids the need to correct for population structure or any genetic background, as the method implemented in the \pkg{netgwas} package simultaneously associates all markers to the phenotype. 
 In contrast, most GWAS methods rely on population structure correction to avoid false genotype-phenotype associations due to their single-loci approach \citep{yu2006unified, kang2008efficient, kang2010variance, lippert2011fast}.

\begin{algorithm}[!b]
	\renewcommand{\algorithmicrequire}{\textbf{Input:} }
	\renewcommand{\algorithmicensure}{\textbf{Output:} }
	\caption{Monte Carlo Gibbs sampling for estimating $\bar{R}$ in (\ref{Estep}) }
	\label{algorithm:gibbs}
	\begin{algorithmic}[1]
		\REQUIRE A data set containing the variables $Y^{(i)}$ for $i = 1 \ldots, n$.
		\ENSURE  Mean of the conditional expectation, $\bar{R} = \frac{1}{n} \sum\limits_{i=1}^{n} E( Z^{(i)} Z^{(i)^T} | y^{(i)};  \mathcal{D}, \Theta^{(m-1)})$ in  (\ref{Estep}).
		\vspace{0.1cm}
		\STATE For each $j = \{1, \ldots,p\} $ generate the latent data from $Y_j= F^{-1}_j(\Phi(Z_j))$, where $F_j$ and $\Phi$ define the empirical marginals and the CDF of standard normal distribution, respectively for $i = 1 \ldots, n$ and $j = 1, \ldots, p$;
		
		\STATE Estimate $\mathcal{D}$, vectors of lower and upper truncation points, whose boundaries come from  $Y^{(i)}$ for $i = 1 \ldots, n$; 
		\FOR {$i = 1, \ldots, n$}
		\STATE Sample from a p-variate truncated normal distribution with the boundaries in Line 2 above; 
		\FOR{$N$ iterations}
		\STATE Estimate $R^{(i), N} = E(Z_\star^{(i)} Z_\star^{(i)T} | y^{(i)}, \widehat{\Theta}^{(m)})$, where $
		Z_\star^{(i)}=
		\left( {\begin{array}{c}
				Z_\star^{(i)1}\\   \vdots    \\ Z_\star^{(i)N} \      \end{array} } \right) \in \mathbb{R}^{N \times p}$;
		\ENDFOR
		\STATE Update $\widehat{R}^{(i)} =  \frac{1}{N} Z_\star^{(i)} Z_\star^{(i)T} $;
		\ENDFOR
		\STATE Calculate  $\widehat{\bar{R}} = \frac{1}{n} \sum\limits_{i=1}^{n} \widehat{R}^{(i)}$.
	\end{algorithmic}
\end{algorithm}

\section{Technical details}
\label{chpter4:method}
Graphical models combine graph theory and probability theory to create networks that model complex probabilistic relationships. Undirected graphical models represent the joint probability distribution of a set of variables via a graph $G=(V,E)$, where $V = \{1, 2, \ldots, p\}$ specifies the set of random variables and $E \subset V\times V$ represents undirected edges $(i,j) \in E \Leftrightarrow (j,i) \in E$. 
The pattern of edges in the graph represents the conditional dependencies between the variables; the absence of an edge between two nodes indicates that any statistical dependency between these two variables is mediated via some other variable or set of variables. The conditional dependencies between variables, which are represented by edges between nodes, are specified via parameterized conditional distributions. 
We refer to the pattern of edges as the structure of the graph. In this paper, the goal is to learn the graph structure from ordinal data and mixed ordinal-and-continuous data.

\paragraph{Sparse latent graphical model.} A p-dimensional copula $\mathcal{C}$ is a multivariate distribution with uniform margins on $[0, 1]$. Any joint distribution function can be written in terms of its marginals and a copula which encodes the dependence structure. Here we consider a subclass of joint distributions encoded by the Gaussian copula 
$F(y_1, \ldots, y_p) = \Phi_p \Big( \Phi^{-1}(F_1(y_1)), \ldots, \Phi^{-1}(F_p(y_p)) \ | \ \Omega \Big)$
where $\Phi_p(. \ | \ \Omega)$ is the cumulative distribution function (CDF) of p-variate Gaussian distribution with correlation matrix $\Omega$; $\Phi$ is the univariate standard normal CDF; and $F_j$ is the CDF of $j$-th variable, $Y_j$ for $j = 1, \ldots, p$. Note that $y_j$ and $y_{j'}$ are independent if and only if $\Omega_{jj'} = 0$. 

A Gaussian copula can be written in terms of latent variables $Z$: Let $F_j^{-1}(y) = \mbox{inf}\{ y: F_j(x) \ge y, x \in \mathcal{R} \}$ be the pseudo-inverse of the marginals. 
Then a Gaussian copula is defined as
$Y_{ij} = F_j^{-1} (\Phi(Z_{ij}))$ where $Z \sim \mathcal{N}_p (0, \Omega)$ and 
$Y= (Y_1, \ldots, Y_p)$ and $Z= (Z_1, \ldots, Z_p)$ represent the non-Gaussian observed variables and Gaussian latent variables, respectively. 
Without lose of generality, we assume that $Z_j$'s have unit variances of $\sigma_{jj}= 1$ for $j = 1,\ldots,p$. Thus, $Z_j$'s marginally follow standard Gaussian distribution. Each observed variable $Y_j$ is discretized from its latent counterpart $Z_j$. For the $j$-th latent	variable ($j = 1, \ldots,p$), we assume that the range $(-\infty, \infty)$ splits into $K_j$ disjointed intervals by a set of thresholds $ -\infty = t^{(0)}_j < t^{(1)}_j < \ldots < t^{(K_j-1)}_j < t^{(K_j)}_j = \infty $ such that $Y_j = k$ if and only if $Z_j$ falls in the interval $(t^{(K-1)}_j , t^{(K)}_j)$. Let the parameter $\mathcal{D} = \{ t^{(k)}_j:  j =1,\ldots,p; k = 1, \ldots, K_j\}$ holds the boundaries for the truncation points
 and $z^{(1:i)} = [z^{(1)}, \ldots, z^{(n)}]$ where  $z^{(i)} = (z_1^{(i)}, \ldots, z_p^{(i)})$. 
In order to learn the graph structures, our objective is to estimate the precision matrix $\Theta = \Omega^{-1}$ from $n$ independent observations $y^{(1:i)} = [y^{(1)}, \ldots, y^{(n)}]$, where $y^{(i)} = (y_1^{(i)}, \ldots, y_p^{(i)})$. The conditional independence between two variables given other variables is equivalent to the corresponding element in the precision matrix being zero, i.e. $\theta_{ij} = 0$; or put another way, a missing edge between two variables in a graph G represents conditional independence between the two variables given all other variables. 

In the classical low-dimensional setting of $p$ smaller than $n$, it is natural to implement a maximum likelihood approach to obtain the inverse of the sample covariance matrix. However, in modern applications the dimension $p$ is routinely far larger than $n$, meaning that the inverse sample covariance matrix does not exist. Motivated by the sparseness assumption of the graph we tackle the high-dimensional inference problem for discrete $Y$'s by a penalized expectation maximization (EM) algorithm as

\begin{eqnarray}
		\centering
		\label{Estep}
		\mbox{Q}(\Theta \ | \ \widehat{\Theta}^{(m-1)} ) = \frac{n}{2} \Big[ \log \det(\Theta) - \mbox{tr}( \bar{R} \Theta ) \Big] 
\end{eqnarray}
and
\begin{eqnarray}
\centering
\label{Mstep}
\widehat{\Theta}^{(m)} = \arg \max_{\Theta} \tilde{\mbox{Q}}(\Theta | \hat{\Theta}^{(m-1)}) - \sum\limits_{j \ne j'}^ p\mbox{P}_\rho(| \theta_{jj'}| ),
\end{eqnarray}
where $m$ is the iteration number within the EM algorithm. 
The last term in equation (\ref{Mstep}) represents different penalty functions. Here we impose the sparsity by means of $L_1$ penalty, on the $jj'$-th element of the precision matrix.

We compute the conditional expectation $\bar{R} = \frac{1}{n} \sum\limits_{i=1}^{n} E\Big( Z^{(i)} Z^{(i)^T} \ | \ y^{(i)}; \mathcal{\widehat{D}}, \widehat{\Theta}^{(m-1)}\Big) $ in equation (\ref{Estep}) using two different approaches: numerically through a Monte Carlo (MC) sampling method as explained in algorithm \ref{algorithm:gibbs}, and through a first order approximation based on algorithm \ref{algorithm:approx}. The most flexible and generally applicable approach for obtaining a sample in each iteration of an MCEM algorithm is through a Markov chain Monte Carlo (MCMC) routine like Gibbs and Metropolis – Hastings samplers \citep{metropolis1953equation, hastings1970monte, geman1984stochastic}. Alternatively, the conditional expectation in equation (\ref{Estep}) can be computed by using an efficient approximation approach which calculates elements of the empirical covariance matrix using the first and second moments of a truncated normal distribution with mean $\mu_{ij} = \widehat{\Omega}_{j, -j} \widehat{\Omega}^{(-1)}_{-j,-j} z^{(i)^T}_{-j}$ and variance $\sigma_{i,j}^2 = 1 - \widehat{\Sigma}_{j,-j} \widehat{\Sigma}^{-1}_{-j,-j} \widehat{\Sigma}_{-j,-j}$ (see \citet{behrouzi2019detecting} for details).
The two proposed approaches are practical when some observations are missing. For example, if genotype information for genotype $j$ is missing, it is still possible to draw Gibbs samples for $Z_j$ or approximate the empirical covariance matrix, as the corresponding conditional distribution is Gaussian.

The optimization problem in (\ref{Mstep}) can be solved efficiently in various ways by using glasso or CLIME approaches \citep{friedman2008sparse, hsieh2011sparse}. Convergence of the EM algorithm for penalized likelihood problems has been proved in \citet{green1990use}. Our experimental study shows that the algorithm usually converges after several iterations $(< 10)$. Note that the sparsity of the estimated precision matrix in Equation (\ref{Mstep}) is controlled by a vector of penalty parameter $\rho$. We follow \cite{foygel2010extended} in using the extended Bayesian information criterion (eBIC) to select a suitable regularization parameter $\rho^*$ to produce a sparse graph with a sparsity pattern corresponding to $\widehat{\Theta}_{\rho^*}$. Alternatively, instead of using the EM algorithm, a nonparanormal skeptic approach can be used to estimate graph structure through Spearman's rho and Kendall's tau statistics; details can be found in \citet{liu2012high}. 



\begin{algorithm}[!b] 
	\renewcommand{\algorithmicrequire}{\textbf{Input:}}
	\renewcommand{\algorithmicensure}{\textbf{Output:}}
	\caption{ Approximation of the conditional expectation in (\ref{Estep})}
	\label{algorithm:approx}
	\begin{algorithmic}[1]
		\REQUIRE A $(n \times p)$ data matrix $Y$, where Y$^{(i)}_j = F^{-1}_j(\Phi(Z^{(i)}_{j}))$ the $F_j$ and $\Phi$ define the empirical marginals and the CDF of standard normal, respectively for $i = 1 \ldots, n$ and $j = 1, \ldots, p$; 
		\ENSURE The conditional expectation $R = \frac{1}{n} \sum\limits_{i=1}^{n} E( Z^{(i)} Z^{(i)^T} | y^{(i)}; \mathcal{\widehat{D}}, \mathbf{\Theta})$; 
		\STATE Initialize $E(z^{(i)}_{j} \ | \ y^{(i)}; \mathcal{\widehat{D}}, \widehat{ \mathbf{\Theta}}) \approx E(z^{(i)}_{j} \ | \ y^{(i)}_{j}; \mathcal{\widehat{D}} )$, $E( (z_j{^{(i)}})^2 \ | \ y^{(i)}; \mathcal{\widehat{D}}, \widehat{ \mathbf{\Theta}}) \approx E( (z_j{^{(i)}})^2 \ | \ y^{(i)}_j; \mathcal{\widehat{D}} )$, and  $E(z^{(i)}_{j} z^{(i)}_{j'} \ | \ y^{(i)}; \mathcal{\widehat{D}}, \widehat{ \mathbf{\Theta}}) \approx  E(z^{(i)}_{j} \ | \ y^{(i)}_{j}; \mathcal{\widehat{D}})  E(z^{(i)}_{j'} \ | \ y^{(i)}_{j'}; \mathcal{\widehat{D}} )$ for $i = 1,\ldots, n$ and $j,j' = 1, \ldots, p$;
		
		\STATE Initialize $r_{j,j'}$ for $1 \leq j, j' \leq p$ using the Line 1 above, then estimate $\widehat{ \mathbf{\Theta}}$ by maximizing (\ref{Mstep});
		\FOR {$i = 1, \ldots, n$} 
		\IF{$j = j'$}
		\STATE Calculate $E((z_j{^{(i)}})^2 \ | \ y^{(i)}_j; \mathcal{\widehat{D}}, \widehat{ \mathbf{\Theta}} )$  for $j = 1, \ldots, p$; 
		\ELSE
		\STATE Calculate $E(z^{(i)}_{j} \ | \ y^{(i)}; \mathcal{\widehat{D}}, \widehat{ \mathbf{\Theta}})$  and then 
		
		$E(z^{(i)}_{j} z^{(i)}_{j'} \ | \ y^{(i)}; \mathcal{\widehat{D}}, \widehat{ \mathbf{\Theta}}) =  E(z^{(i)}_{j} \ | \ y^{(i)}; \mathcal{\widehat{D}}, \widehat{ \mathbf{\Theta}})  E(z^{(i)}_{j'} \ | \ y^{(i)}; \mathcal{\widehat{D}}, \widehat{ \mathbf{\Theta}} )$ for $i = 1,\ldots, n$ and $j = 1, \ldots, p$;
		\ENDIF
		\ENDFOR
		\STATE Calculate $r_{j,j'}= \frac{1}{n} \sum_{i=1}^{n} E(z_j^{(i)} z_j^{(i)t} \ | \ y^{(i)}; \mathbf{\widehat{\mathcal{D}}}, \widehat{ \mathbf{\Theta}} )$ for $ 1 \le j = j' \le p$.
	\end{algorithmic}
\end{algorithm}

\subsection{Extension to linkage map construction}
Here we convert the estimated network to a one-dimensional map using two different approaches. Depending on the type of (experimental) population (i.e. inbred or outbred), we order markers based on dimensionality reduction or based on bandwidth reduction, which both result in an one-dimensional map.

In inbred populations, loci in the genome of the progenies can be assigned to their parental homologues, resulting in a simpler conditional independence relationship between neighboring markers. Here, we use multidimensional scaling (MDS) to project markers in a p-dimensional space onto a one-dimensional map while attempting to maintain pairwise distances. Let $G(V^{(d)},E^{(d)})$ be a sub--graph on the set of unordered $d$ markers, where $V^{(d)} = \{1, \ldots, d\}$, $d \leq p$ and the edge set $E^{(d)}$ represents all the links among $d$ markers. We calculate the distance matrix $D$ as follows
\begin{eqnarray}
D_{ij} = \left\{ 
\begin{array}{ll}
- \log (r_{ij}) & \mbox{if $i\ne j$}\\
0  & \mbox{if $i = j$},
\end{array} \right.
\end{eqnarray}
\begin{eqnarray}
r_{ij} = - \frac{\theta_{ij}}{\sqrt{\theta_{ii}\theta_{jj}}},
\end{eqnarray}
where $\theta_{ij}$ is the $ij$-th element of the precision matrix  $\widehat{\Theta}_{\rho^*}$. We aim to construct a configuration of $d$ data points in a one--dimensional Euclidean space by using information about the distances between the $d$ nodes. In this regards, we define a sequential ordering $L$ of $d$ elements such that the distance $\widehat{D}$ between them is similar to $\mathcal{D}$. We consider a metric multi-dimensional scaling
\begin{eqnarray}
\widehat{E}= \mbox{arg}\min_L \sum\limits_{i=1}^{d}\sum\limits_{j=1}^{d}(D_{ij} - \widehat{D}_{ij})^2,
\end{eqnarray}
that minimizes the so called mapping error $\widehat{E}$ across all sequential orderings \citep{sammon1969nonlinear}.

We propose a different ordering algorithm for outbred populations. In these populations, mating of two non-homozygous parents result in markers in the genome of progenies that cannot easily be mapped into their parental homologues. 
To order markers in outbred populations, we use the Cuthill-McKee (RCM) algorithm \citep{cuthill1969reducing} to permute the sparse matrix  $\widehat{\Theta}_\rho^{(d)}$ that has a symmetric sparsity pattern into a band matrix form with a small bandwidth. The bandwidth of the associated adjacency matrix $A$ is defined as $\beta = \max_ {A_{ij} \ne 0} | i - j|$. The algorithm produces a permutation matrix $P$ such that $P A P^T$ has a smaller bandwidth than matrix $A$ does. The bandwidth decreases by moving the non-zero elements of the matrix $A$ closer to the main diagonal. The way to move the non-zero elements is determined by relabeling the nodes in graph $G(V_d, E_d)$ in consecutive order. Moreover, all of the nonzero elements are clustered near the main diagonal.

\section{Package design and functionality}
\label{interface}
The \pkg{netgwas} R package implements the Gaussian copula graphical models \citep{behrouzi2019detecting} for (i) constructing linkage maps in diploid and polyploid species and learning (ii) linkage disequilibrium networks and (iii) genotype-phenotype networks. Below, we illustrate the three main functions using a diploid A.thaliana population, a tetraploid potato, and maize NAM populations. Given that the computational cost for the usual size of GWAS data $( > 10^5)$ is expensive, we use small data sets to explain the functionality of the package. All the results can be replicated using the functions in the \pkg{netgwas} package (see Supplementary Materials).

\subsection{Linkage map construction}
\label{map}
This module reconstructs linkage maps for diploid and polyploid organisms. Diploid organisms contain two copies of each chromosome, one from each parent, whereas polyploids contain more than two copies of each chromosome. In polyploids the number of chromosome sets reflects their level of ploidy: triploids have three copies, tetraploids have four, pentaploids have five, and so forth. 

Typically, mating is between two parental lines that have recent common biological ancestors; this is called inbreeding. If they have no common ancestors up to roughly $4$-$6$ generations, then this is called outcrossing. In both cases the genomes of the derived progenies are random mosaics of the genome of the parents. However, in the case of inbreeding parental alleles are distinguishable in the genome of the progeny, in outcrossing this does not hold.   

Some \emph{inbreeding} designs, such as Doubled haploid (DH), lead to a homozygous population where the derived genotype data include only homozygous genotypes of the parents namely AA and aa (conveniently coded as $0$ and $1$) for diploid species. Other inbreeding designs, such as F2, lead to a heterozygous population where the derived genotype data contain heterozygous genotypes as well as homozygous ones, namely aa, Aa, and AA for diploid species, conveniently coded as $0$, $1$ and $2$ which correspond to the dosage of the reference allele A. We remark that the Gaussian copula graphical models help us to keep heterozygous markers in the linkage map construction, rather than turn them to missing values as most other methods do in map construction for recombinant inbred line (RIL) populations.

\emph{Outcrossing} or outbred experimental designs, such as full-sib families, derive from two non-homozygous parents. Thus, the genome of the progenies includes a mixed set of many different marker types containing fully informative markers 
and partially informative markers 
. Markers are called fully informative when all of the resulting gamete types can be phenotypically distinguished on the basis of their genotypes; markers are called partially informative when they have identical phenotypes \citep{wu2002simultaneous}.

\subsubsection*{ {\tt netmap()}}
The {\tt netmap()} function handles various inbred and outbred mapping populations, including recombinant inbred lines (RILs), F2, backcross, doubled haploid, and full-sib families, among others. Not all existing methods for linkage mapping support all inbreeding and outbreeding experimental designs. However, our proposed algorithm constructs a linkage map for any type of experimental design of biparental inbreeding and outbreeding.  
Also, it covers a wide range of possible population types. Argument {\tt cross} in the map function must be specified to choose an ordering method. In inbred populations, markers in the genome of the progenies can be assigned to their parental homologous, resulting in a simpler conditional independence pattern between neighboring markers. In the case of inbreeding, we use multidimensional scaling (MDS). A metric MDS is a classical approach that maps the original high-dimensional space to a lower dimensional space, while attempting to maintain pairwise distances. An outbred population derived from mating two non-homozygous parents results in markers in the genome of progenies that cannot be easily assigned to their parental homologues. Neighboring markers that vary only on different haploids will appear as independent, therefore requiring a different ordering algorithm. In that case, we use the reverse Cuthill-McKee (RCM) algorithm \citep{cuthill1969reducing} to order markers.

The function can be called with the following arguments
\begin{verbatim}
	netmap(data, method = "npn", cross = NULL, rho = NULL, n.rho = NULL, rho.ratio = NULL, 
	       min.m = NULL, use.comu = FALSE, ncores = "all", verbose = TRUE)
\end{verbatim}
The main task of this function is to construct a linkage map based on conditional (in)dependence relationships between markers, which can be estimated using the methods, {\tt ``gibbs"}, {\tt``approx"}, and {\tt``npn"}. The estimation procedure relies on maximum penalized log-likelihood, where the argument {\tt rho}, a decreasing sequence of non-negative numbers, controls the sparsity levels, which corresponds to the last term in Equation (\ref{Mstep}). Leaving the input as {\tt rho = NULL}, the program automatically computes a sequence of {\tt rho} based on {\tt n.rho} and {\tt rho.ratio}. The argument {\tt n.rho} specifies the number of regularization parameters (the default is 6) and {\tt rho.ratio} determines the ratio between the consecutive elements of {\tt rho}. Depending on the population type, {\tt inbred} or {\tt outbred}, different algorithms are applied to order markers in the genome. If it is known, the user can specify an expected minimum number of markers in a linkage group (LG) via the argument {\tt min.m}. Furthermore, linkage groups can be identified either using the fast greedy community detection algorithm \citep{newman2004fast} or simply each disconnect sub-networks can form a linkage group. The {\tt ncores = "all"} automatically detects number of available cores and runs the computations in parallel on (available cores -1).

The {\tt netmap()} function returns an object of the {\tt S3} class type {\tt netgwasmap} and {\tt plot.netgwasmap} and {\tt print.netgwasmap} are summary method functions for this object class. 
The {\tt netgwasmap} mainly holds the estimated linkage map (in object {\tt map}) and a list containing all output results (in object {\tt res}) of the regularization path {\tt rho}.

\subsubsection{\textbf{{\tt buildMap()}}}
The function {\tt buildMap()} allows users to interact with the map construction procedure and to build the linkage map on the manually selected penalty term. Whereas the function {\tt netmap()} selects the optimal penalty term $\rho^\star$ using the eBIC method.

The function can be called via
\begin{verbatim}
	buildMap(res, opt.index, min.m = NULL, use.comu = FALSE).
\end{verbatim}

The argument {\tt opt.index} can be chosen manually which is a number between one and the number of penalty parameter {\tt n.rho} in {\tt netmap()}. In the default setting, the {\tt n.rho} is $6$. So, the {\tt opt.index} can get a value between $1$ and $6$. Like function {\tt netmap()}, the argument {\tt min.m} is an optional argument in {\tt buildMap()} function, where it keeps the clusters of markers that at least have a size of {\tt min.m} member of markers. The default value for this argument is 2.
The {\tt use.comu} argument is an alternative approach to find linkage groups. 
The {\tt use.comu} argument is an alternative approach to find linkage groups. 

\begin{figure*}[t]
	\centering
	\includegraphics[width=0.45\textwidth]{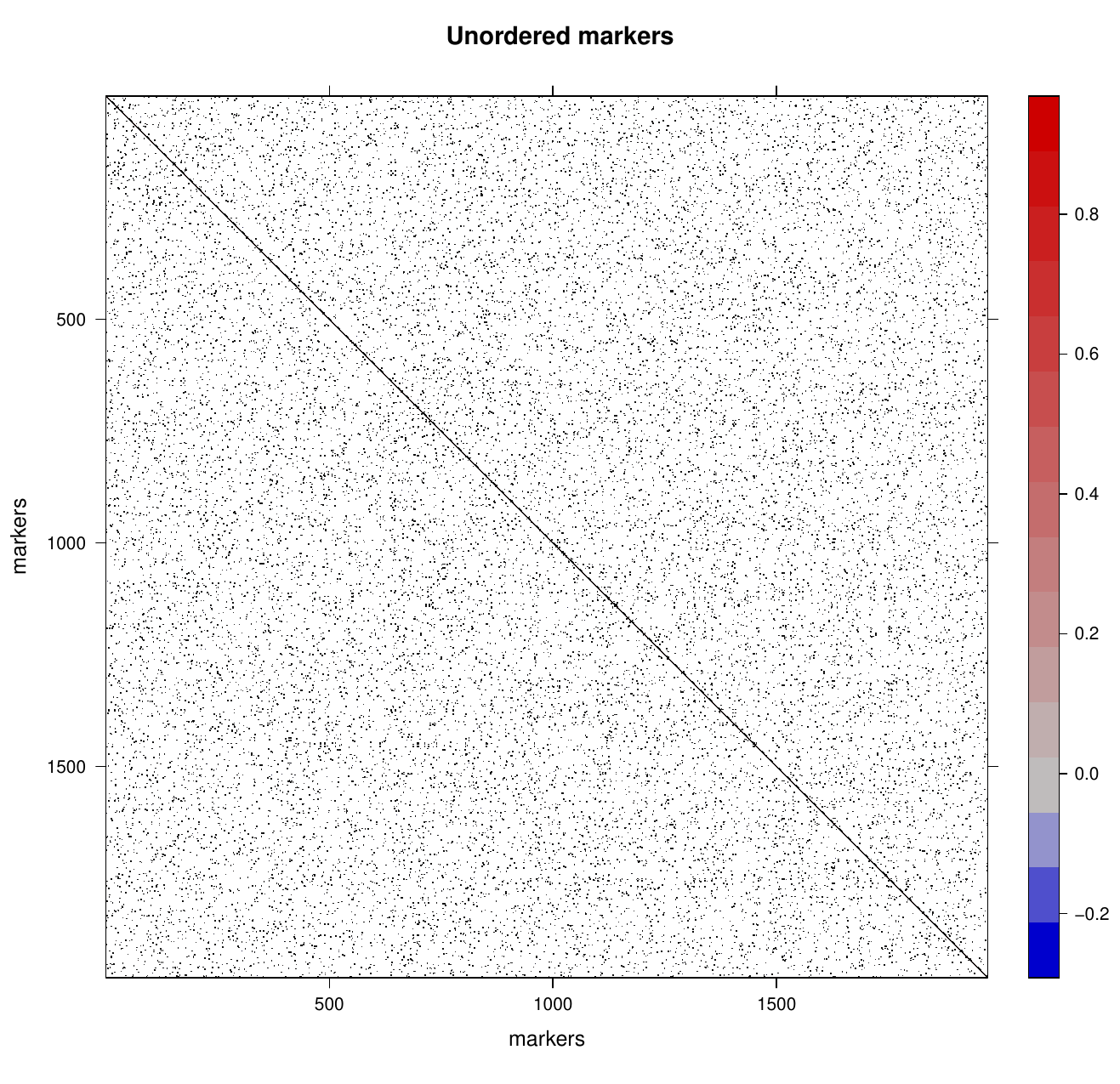} \hspace{0.5cm}
	\includegraphics[width=0.45\textwidth]{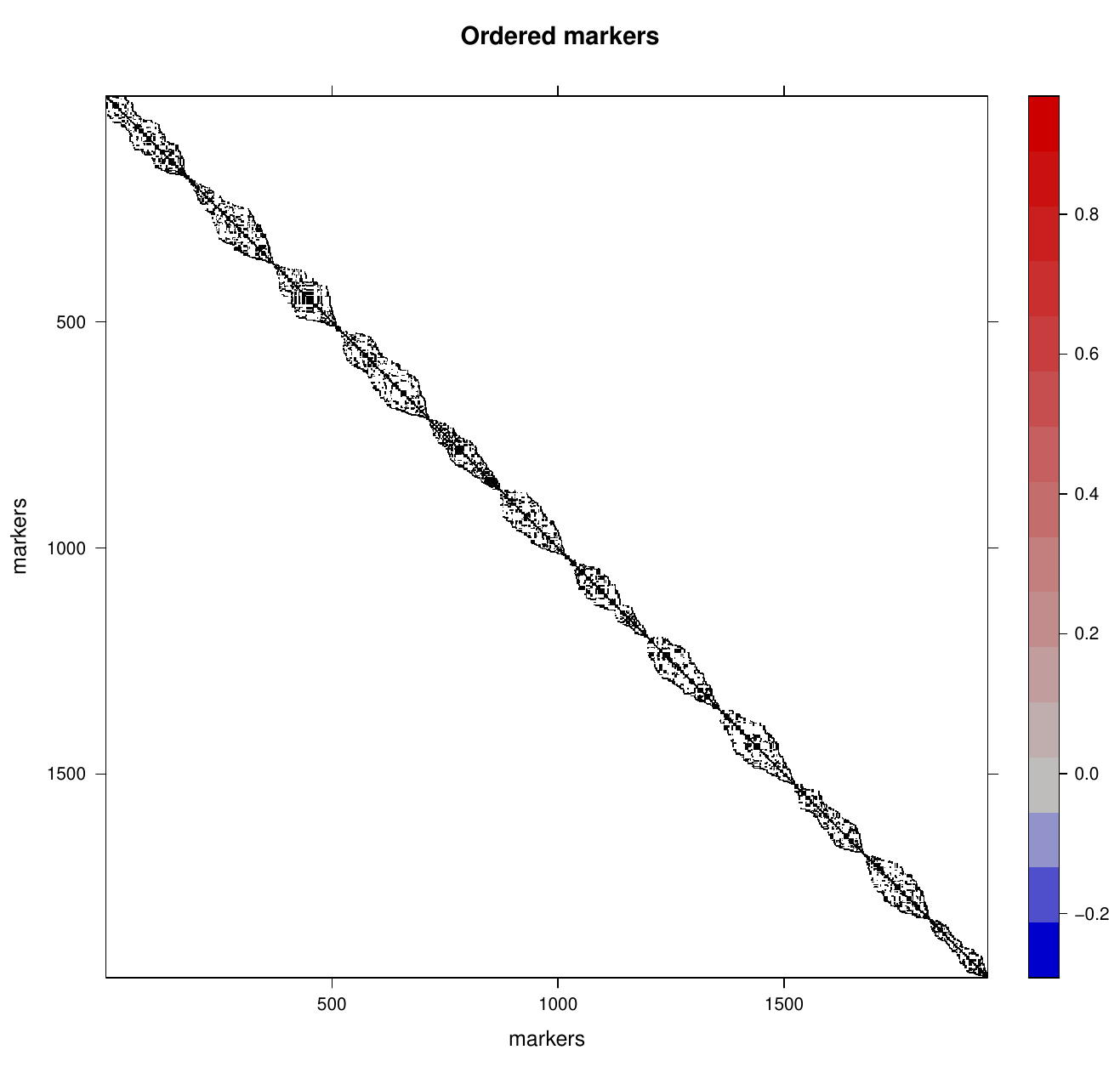} 
	
	(a) \hspace{6.5cm}  (b)
	\vspace{-0.3cm}
	\caption{Linkage map construction in tetraploid potato. A total of $156$ F1 plants were genotyped across $1972$ genetic markers. (a) The estimated partial correlation matrix for the genotyping data. (b) The estimated partial correlation matrix after ordering the genetic markers, where 
		all the 12 potato chromosomes 
		are detected correctly.} 
	\label{potato2015}
\end{figure*}

\subsection{Detecting linkage disequilibrium networks}
\label{marker-marker network}
The function \texttt netsnp() reconstructs a high-dimensional linkage disequilibrium interactions network for diploid and polyploid (GWAS) genotype data. Genetic viability can be considered as a phenotype. This function detects the conditional dependent short- and long-range linkage disequilibrium structure of genomes and thus reveals aberrant marker-marker associations that are due to epistatic selection. In other words, this function detects intra-- and inter--chromosomal conditional interactions networks and can be called via
\begin{verbatim}
	netsnp(data, method = "gibbs", rho = NULL, n.rho = NULL, rho.ratio = NULL,
	       ncores = "all", verbose = TRUE)
\end{verbatim}
for any bi-parental genotype data containing at least two genotype states and possibly missing values. The input data can be either an (n $\times$ p) genotype data matrix, an object of class {\tt netgwasmap}, which is an output of functions {\tt netmap()} and {\tt buildMap()}, or a simulated data from the function {\tt simgeno()}. Depending on the dimension of the input data a suitable `{\tt method}' and its related arguments can be specified. The argument {\tt ncores} determines the number of cores to use for the calculations. Using {\tt ncores = "all"} automatically detects number of available cores and runs the computations in parallel.

This function returns an $S3$ object of class "netgwas", which holds mainly the following objects: (i) {\tt Theta} a list of estimated $(p \times p)$ precision (inverse of variance-covariance) matrices that infer the conditional independence relationships patterns among genetic loci, (ii) {\tt path} which is a list of estimated  $(p \times p)$ adjacency matrices. This is the graph path corresponding to {\tt Theta}, (iii) {\tt rho} which is a vector with {\tt n.rho} dimension containing the penalties, and (iv) {\tt loglik} contains the maximum log-likelihood values along the graph path. To select an optimal graph the function {\tt selectnet()} can be used.

\subsection{Reconstructing genotype-phenotype networks}
\label{QTL}
Complex genetic traits are influenced by multiple interacting loci, each with a possibly small effect. Our approach reduces the number of candidate genes from hundreds to much fewer genes. It is of great interest to geneticists and biologist to discover a set of most effective genes that directly affect a complex trait in GWAS. To overcome the limitations of traditional analysis, such as single-locus association analysis (looking for main effects of single marker loci), multiple testing and QTL analysis, we use the proposed mixed graphical model to study the simultaneous associations between phenotypes and SNPs. Our method allows for a more accurate interpretation of findings, because it adjusts for the effects of remaining variables --SNPs and phenotypes-- while measuring the pairwise associations, whereas the traditional methods use marginal associations to often analyze SNPs and phenotypes one at a time .

Graphical modeling is a powerful tool for describing complex interaction patterns  among variables in high-dimensional data used frequently in microarray analysis \citep{butte2000discovering}. 
In our modelling framework, a ge\-no\-type--phenotype network is a complex network made up of interactions among: (i) genetic markers, (ii) phenotypes (e.g. disease), and (iii) between genetic markers and phenotypes. The first problem in analyzing genotype-phenotype data is the mixed variable-types, where markers are ordinal (counting the number of a major allele), and phenotypes (disease) can be measured in continuous or discrete scales. We deal with mixed  discrete-and-continuous variables by means of copula. A second issue relates to the high-dimensional setting of the data, where thousands of genetic markers are measured across a few samples; we deal with inferring potentially large networks with only few biological samples. Fortunately, genotype-phenotype networks are intrinsically sparse, in the sense that only few elements interact with each other. This sparsity assumption is incorporated into our algorithm based on penalized graphical models. The proposed method is implemented in the  {\texttt netphenogeno()} function, where the input data can be an ($n \times p$) {\tt matrix} or a {\tt data.frame} where $n$ is the sample size and $p$ is the dimension that includes marker data and phenotype measurements. One may consider including more columns, like environmental variables. 

The function is defined by
\begin{verbatim}
	netphenogeno(data, method = "npn", rho = NULL, n.rho = NULL, rho.ratio = NULL, 
	               ncores = "all", em.iter = 5, em.tol = .001, verbose = TRUE)
\end{verbatim}
and reconstructs genotype-phenotype interactions network for an input data of a genotype-phenotype data matrix or a data.frame. Detecting interactions network among genotypes, phenotypes and environmental factors is also possible using this function. Depending on the size of the input data, the user may choose "gibbs", "approx", or "npn" method for learning the networks. For a medium ( $\sim$500) and a large number of variables we recommend to choose "gibbs" and "approx", respectively. Choosing "npn" for a very large number of variables ( >2000) is computationally efficient. The default method is set to "npn". Like the function {\tt netsnp()}, the {\tt netphenogeno()} function returns an object of class \pkg{netgwas}. 

For objects of type  `netgwas' there are plot, print and summary methods available. The plotting function {\tt plot.netgwas()} provides a visualization plot to monitor the path of estimated networks for a range of penalty terms. The functions {\tt plot.netgwasmap(), {\tt plot.select()} and {\tt plot.simgeno()} } visualize the corresponding network, the optimal graph and the results of model-based simulated data, respectively. 

To speed up computations in all the three key functions of the \pkg{netgwas} package, we use the \pkg{parallel} package on the Comprehensive {\tt R} Archive Network (CRAN) at \url{http://CRAN.R-project.org} to support parallel computing on a multi-core machine to deal with large inference problems. For the optimizing the memory usage, we use the \pkg{Matrix} package \citep{Matrix} for sparse matrix output when estimating and storing full regularization paths for large datasets. The use of these libraries significantly improves the computational speed of the functions within the package.

\begin{figure}[t]
	\centering
	\includegraphics[width=0.7\textwidth]{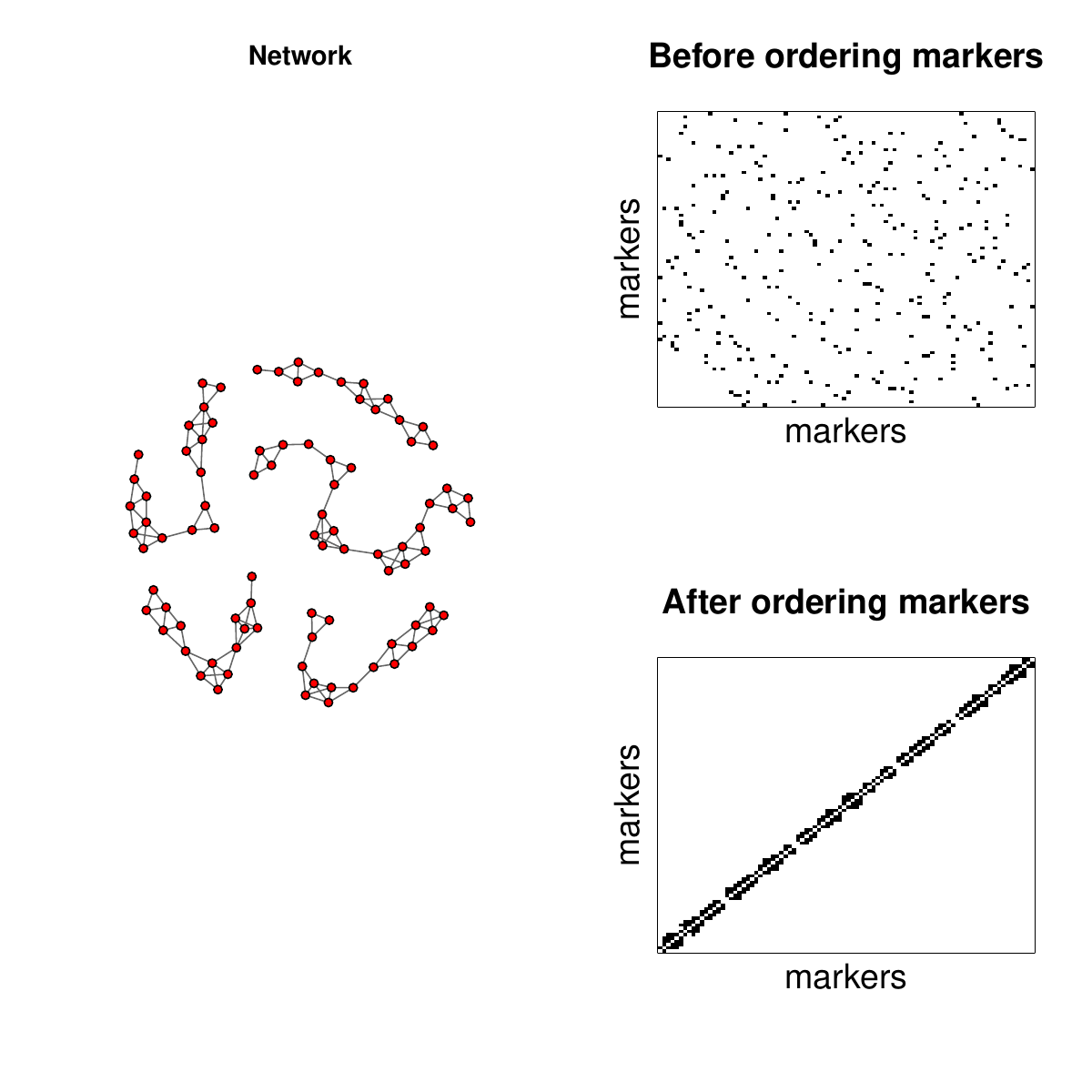} 
	\caption{Linkage map construction in \emph{A.thaliana} using RILs (recombinant inbred lines) population of 367 individuals 
		and 90 genetic markers. The inferred network detects the five chromosomes of A.thaliana and the structure of associations among SNPs (single nucleotide polymorphisms) within chromosomes. The left figures show the partial correlation matrix before and after ordering the SNPs.} 
	\label{plotmpcvicol}
\end{figure}

\section{Application to real datasets}
\label{pkg:netgwas}

\subsection{Linkage map construction}
\label{Example_map}
In Fig \ref{potato2015} and Fig \ref{plotmpcvicol}, we provide two examples output of building linkage map in outbred tetraploid \emph{potato} and inbred diploid \emph{A.thaliana} datasets. The map construction was computed in about 7 minutes for tetraploid potato data and in 0.6 seconds for A.thaliana data on an Intel i7 laptop with 16 GB RAM. 

\paragraph{Linkage map construction in potato.} For the sake of illustration, below we show the steps to construct a linkage map for {\tt TetraPotato} in \pkg{netgwas}. The tetraploid \emph{potato} data are derived from a cross between “Jacqueline Lee” and “MSG227-2”, where $156$ F1 plants were genotyped across $1972$ genetic markers \citep{massa2015genetic}. Five allele dosages are possible in this full-sib autotetraploid mapping population (AAAA, AAAB, AABB, ABBB, BBBB), where the genotypes are coded as $\{0, 1, 2, 3, 4\}$.  This dataset includes $0.07\%$ missing observations. 
\begin{verbatim}
data(tetraPotato)
# Shuffle the order of markers
dat <- tetraPotato[ , sample(ncol(tetraPotato))] 
potato.map <- netmap(dat, cross = "outbred")
potato.map.ordered <- buildMap(potato.map, opt.index = 3)
potato.map.ordered
\end{verbatim}
{\footnotesize
	\begin{verbatim}
	Number of linkage groups:  12 
	Number of markers per linkage group: 165 157 129 153 183 196 173 148 152 161 187 146 
	Total number of markers in the linkage map: 1950.(22  markers removed from the input genotype data)
	Number of sample size: n = 156 
	Number of categories in dataset: 5  ( 0 1 2 3 4 ) 
	The estimated linkage map is inserted in <OUTPUT NAME>$map 
	To visualize the network consider plot(<OUTPUT NAME>) 
	----------------------- 
	To visualize the other associated networks consider plot(<OUTPUT NAME>$allres) 
	\end{verbatim}
}
\begin{verbatim}

plot(potato.map.ordered, vis = "unordered markers") 
plot(potato.map.ordered, vis = "ordered markers") 
map <- potato.map.ordered$map
\end{verbatim}

The argument {\tt vis} in the above {\tt plot} function can be fixed to {\tt "interactive"}, which it gives a better network resolution particularly for a large number of nodes. Fig \ref{potato2015} visualizes a summary of its mapping process, where Fig \ref{potato2015}a shows the conditional dependence pattern between unordered SNP markers in the Jacqueline Lee $\times$ MSG227-2 population. Fig \ref{potato2015}b shows the structure of the selected graph after ordering markers. All 12 potato chromosomes were detected correctly. 
The tetraploid potato map construction was computed in about 7 minutes on an Intel i7 laptop with 16 GB RAM. 

\paragraph{Linkage map construction in \emph{A.thaliana.}}  In this example, we construct a linkage map for the \emph{Arabadopsis thaliana} data which are derived from a RIL cross between Columbia-0 (Col-0) and Cape Verde Island (Cvi-0), where $367$ individual plants were genotyped across $90$ genetic markers \citep{simon2008qtl}. The dataset {\tt CviCol} contains 0.2\% missing values and three possible genotype states, where A and B denote parental homozygous loci, coded as 0 and 2, respectively and H denotes heterozygous loci which coded as 1.

\begin{verbatim}
data(CviCol)
set.seed(1)
cvicol <- CviCol[ ,sample(ncol(CviCol))]
out <- netmap(cvicol, cross= "inbred", ncores= 1)
out$opt.index
[1] 6
\end{verbatim}
In the above code, the {\tt out\$opt.index} shows the index of the selected penalty term using the eBIC method. If one is interested in building linkage map, for instance, on the 4th estimated network then the {\tt buildMap()} function can be used as follow
\begin{verbatim}
bm.thaliana <- buildMap(out, opt.index= 4)
bm.thaliana
\end{verbatim}
{\footnotesize
	\begin{verbatim}
	Number of linkage groups:  5 
	Number of markers per linkage group:  24 14 17 16 19 
	Total number of markers in the linkage map: 90. 
	(0 markers removed from the input genotype data) 
	Number of sample size: n = 367 
	Number of categories in dataset: 3  ( 0 1 2 ) 
	The estimated linkage map is inserted in <OUTPUT NAME>$map 
	To visualize the network consider plot(<OUTPUT NAME>) 
	----------------------- 
	To visualize the other associated networks consider plot(<OUTPUT NAME>$allres) 
	To build a linkage map for your desired network consider buildMap() function 
	\end{verbatim}}
\begin{verbatim}
thalianaMap <- bm.thaliana$map
plot(bm.thaliana, vis= "summary")
\end{verbatim}
The estimated linkage map in Fig \ref{plotmpcvicol} is consistent with the existing linkage map in \emph{A.thaliana} \citep{simon2008qtl, behrouzi2019novo}. 

If required, {\tt detect.err()} function detects genotyping errors. This function calculates the error LOD score for each individual at each marker using \cite{lincoln1992systematic} approach; large scores show likely genotyping errors. Here, the \pkg{qtl} package \citep{broman2003r} is used for identification of genotyping errors, where the output gives a list of genotypes that might be in error, when the error LOD scores are smaller than $4$ they can probably be ignored \citep{broman2009brief}. This function supports doubled haploid (DH), backcross (BC), non-advanced recombinant inbred line population with n generations of selfing (RILn) and advanced RIL (ARIL) population types.  

The {\tt cal.pos()} function calculates the genetic distance for diploid populations. It uses the \pkg{qtl} package to calculate genetic distance using different distance functions. 
The {\tt netgwas2cross()} function converts the map object to a {\tt cross} object from \pkg{qtl} package, and vice versa using the function {\tt cross2netgwas()}. These two functions make \pkg{netgwas} flexible with respect to further genetic investigation using \pkg{qtl} package. Furthermore, {\tt cross} objects from the \pkg{qtl} package can also be analyzed using \pkg{netgwas} package.

\begin{figure*}[t]%
	\hspace{-0.3cm}
	\includegraphics[width=0.55\textwidth]{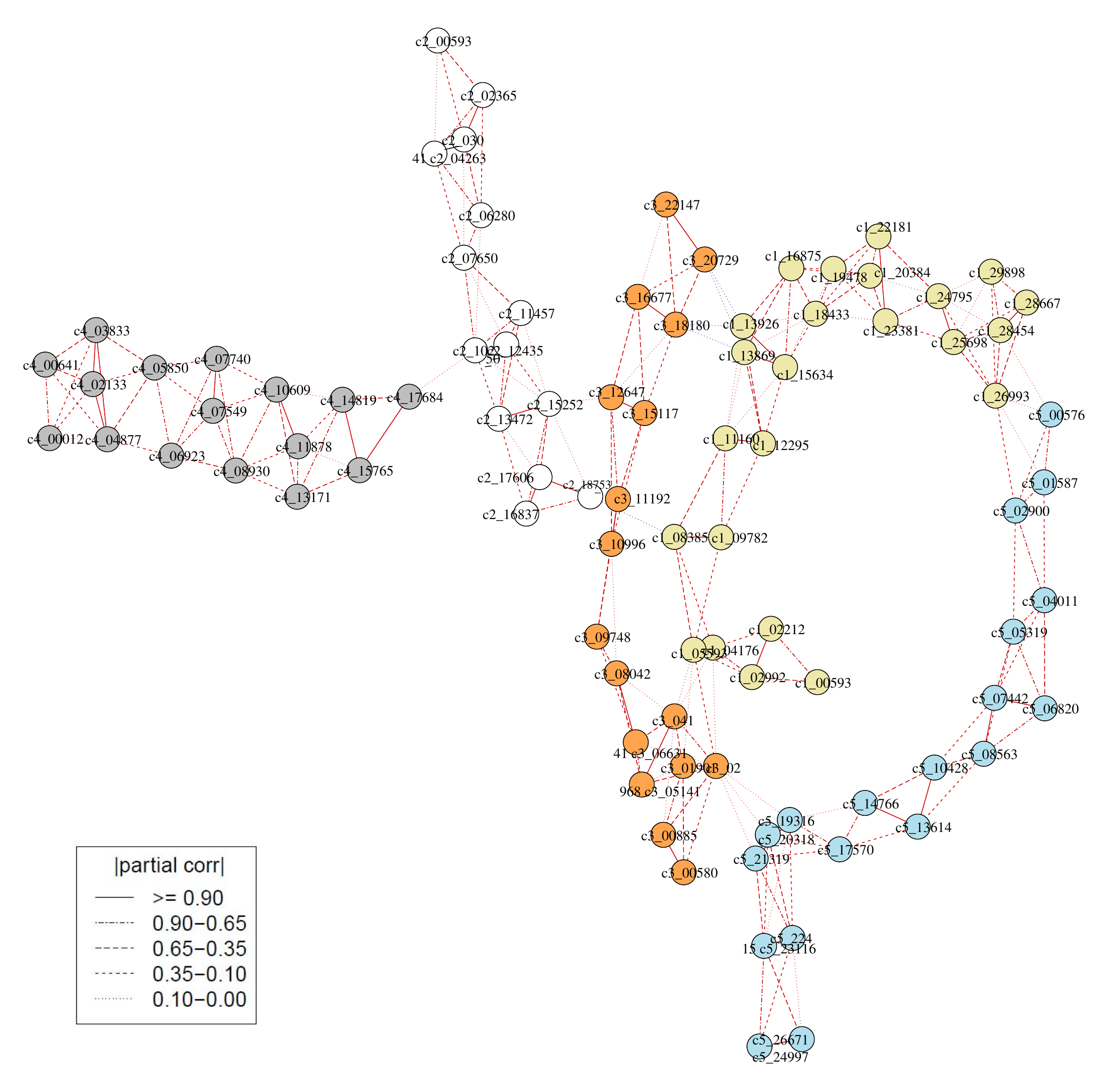}  \hspace{0.1cm} 
	\includegraphics[width=0.45\textwidth]{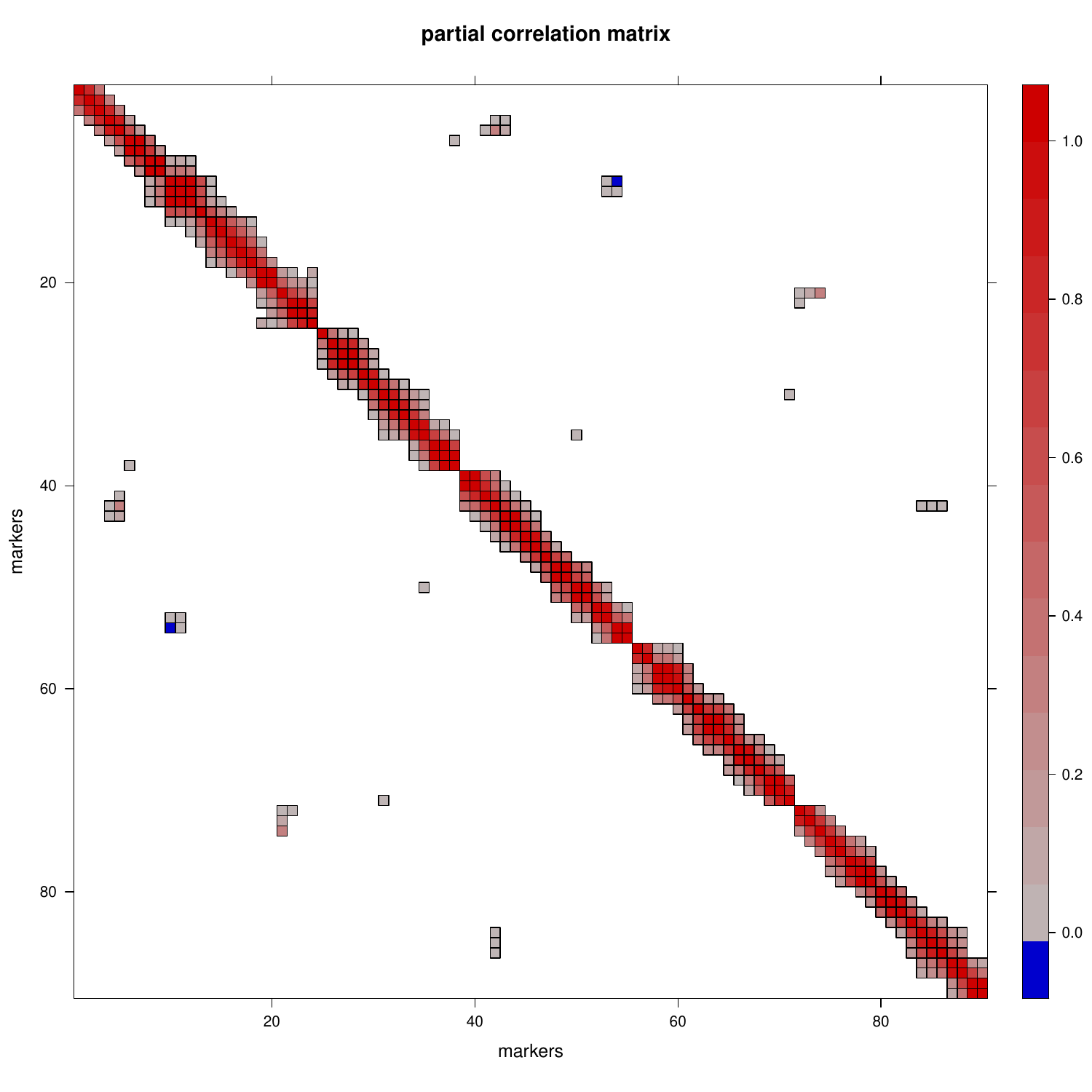}
	
	\hspace{5cm} 	(a) \hspace{5cm}(b)
	\vspace*{-0.1cm}
	\caption{Short- and long-range linkage disequilibrium networks between $90$ markers across the A.\emph{thaliana} genome. (a) Each color corresponds to a different chromosome: yellow, white, orange, gray, and blue represent chromosomes 1 to 5, respectively. Different edge colors show positive {\color{red} \rule{0.5cm}{0.5mm}} and negative {\color{blue} \rule{0.5cm}{0.5mm}} values of partial correlations. (b) Plots simultaneous marker-marker interactions across the genome. Values represent partial correlations.}
	\label{thaliana-intrainter}
\end{figure*}

\subsection{Genome wide association studies}

\paragraph{Linkage disequilibrium networks in \emph{A.thaliana}.} We use the dataset {\tt CviCol} to learn conditionally dependent short- and long-range LD structure in \emph{A.thaliana} genome. The aim here is to identify associations between distant markers that are due to epistatic selection rather than gametic linkage.

\begin{verbatim}
data(CviCol)
set.seed(2)
out <- netsnp(CviCol)
sel <- selectnet(out)
# Steps to visualize the selected network
cl <- c(rep("palegoldenrod", 24), rep("white",14), rep("tan1",17), 
rep("gray",16), rep("lightblue2",19))
plot(sel, vis= "parcor.network", sign.edg = TRUE, layout = NULL, vertex.color = cl)
plot(sel, vis= "image.parcorMatrix", xlab="markers", ylab="markers")
\end{verbatim}

In Fig \ref{thaliana-intrainter}, our method finds that in $Cvi \times Col$ population some trans-chromosomal regions conditionally interact. In particular, the bottom of chromosome 1 and the top of chromosome 5 do not segregate independently of each other. Besides this, interactions between the tops of chromosomes 1 and 3 involve pairs of loci that also do not segregate independently. \cite{bikard2009divergent} studied this genotype data extensively in their lab. They reported that the first interaction (between chr 1 and 5) that our method finds causes arrested embryo development, resulting in seed abortion, and the latter interaction (between chr 1 and 3) causes root growth impairment. In addition to these two regions, we have discovered a few other trans-chromosomal interactions in the A.\emph{thaliana} genome. In particular, two adjacent markers, c1-13869 and c1-13926 in the middle of the chromosome 1, interact epistatically with the adjacent markers, c3-18180 and c3-20729, at the bottom of chromosome 3. The sign of their conditional correlation score is negative, indicating strong negative epistatic selection in $F_2$ population. These markers therefore seem evolutionarily favored to come from the two different $F_0$ grandparents. This suggests some positive effect of the interbreeding of the two parental lines: it could be that the paternal-maternal combination at these two loci protects against some underlying disorder, or that it actively enhances the fitness of the resulting progeny. Regarding the computational time, this example was run in 4 minutes on an Intel i7 laptop with 16 GB RAM.
\begin{figure}[!t]
	\centering
	\includegraphics[width=0.7 \textwidth]{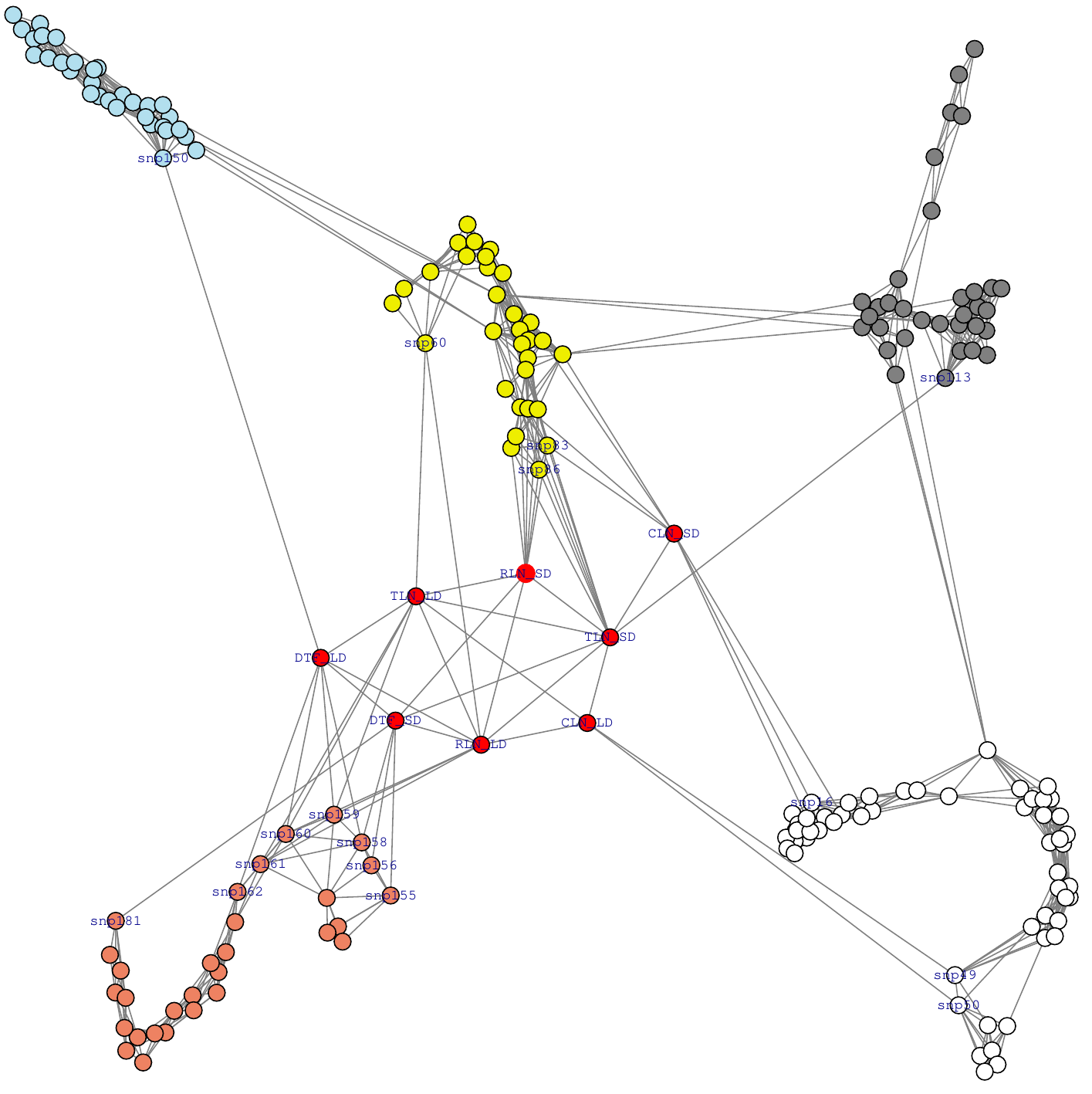}
	\caption{Genotype--phenotype association networks of A.\emph{thaliana}. This shows an example of multi-loci multi-trait genome-wide association analysis. Red nodes show phenotypes; white, yellow, gray, blue, and brown colors stand for chromosomes 1 to 5, respectively. Phenotypes measured in long days (TLN-LD, RLN-LD, DTF-LD) conditionally dependent on a region on top of chromosome 5. 
	Phenotypes measured in short days (CLN-SD, RLN-SD, DTF-SD) are linked mostly to chromosomes 1, 2, and 5. }
	\label{thaliana-phenoGeno}
\end{figure}

\paragraph{Genotype-phenotype networks in \emph{A.thaliana}.} We apply our algorithm to the model plant \emph{Arabidopsis thaliana} dataset, where the accession Kend-L (Kendalville-Lehle; Lehle-WT-16-03) is crossed with the common lab strain Col (Co\-lum\-bi\-a) \citep{balasubramanian2009qtl}. The resulting lines were taken through six rounds of selfing without any intentional selection. The resulting 282 KendC (Kend-L $\times$ Col) lines were genotyped at 181 markers.  Flowering time was measured for 197 lines of this population both in long days, which promote rapid flowering in many A. \emph{thaliana} strains, and in short days. Flowering time was measured using days to flowering (DTF) as well as the total number of leaves (TLN), partitioned into rosette and cauline leaves. In total, eight phenotypes were measured, namely days to flowering (DTF), cauline leaf number (CLN), rosette leaf number (RLN), and total leaf number (TLN) in long days (LD), and DTF, CLN, RLN, and TLN in short days (SD). Thus, the final dataset consists of 197 observations for 189 variables (8 phenotypes and 181 genotypes - SNP markers). 
\begin{verbatim}
data(thaliana)
head(thaliana, n = 3)
     DTF_LD CLN_LD ... DTF_SD CLN_SD RLN_SD TLN_SD snp1  ... snp181
[1,] 17.58   3.42  ...  56.92  12.42  50.92  63.33    2  ...    2
[2,] 17.00   2.58  ...  53.33   8.42  41.58  50.00    0  ...    2
[3,] 27.50   8.08  ...  69.17  15.17  66.92  82.08    2  ...    0

set.seed(12)
out <- netphenogeno(thaliana)
sel <- selectnet(out)

# Steps to visualize the network
cl <- c(rep("red", 8), rep("white",56), rep("yellow2",31),
           rep("gray",33), rep("lightblue2",31), rep("salmon2",30))

id <- c("DTF_LD","CLN_LD","RLN_LD","TLN_LD","DTF_SD","CLN_SD", 
        "RLN_SD", "TLN_SD","snp16", "snp49","snp50", "snp60","snp83", 
         "snp86", "snp113","snp150", "snp155","snp159","snp156",
         "snp161","snp158", "snp160","snp162", "snp181")

plot(sel, vis= "interactive", n.mem= c(8,56,31,33,31,30),
       vertex.color= cl, label.vertex= "some", sel.nod.label= id,
       edge.color= "gray", w.btw= 200, w.within= 20, tk.width = 900,
       tk.height = 900)
\end{verbatim}
The \emph{A.thaliana} genotype-phenotype network in Fig \ref{thaliana-phenoGeno} reveals those SNP markers that are directly affect flowering phenotypes. For example, markers $snp158$, $snp159$, $snp160$, and $snp162$ on chromosome 5 with the assay IDs $44607857$, $44606159$, $44607242$, and $44607209$ regulate the phenotype days to flowering (DTF-LD). For the same phenotype, \cite{balasubramanian2009qtl} have reported a wider range of markers (from $snp158$ to $snp162$ with the assay ID $44607857$ to $44607209$) that associate with DTF-LD. Our obtained smaller markers set is the result of controlling for all possible effects. In particular, the proposed method finds that $snp161$ does not show any association with DTF-LD after adjustments, but $snp159$, $snp160$ and $snp162$ on chromosome $5$ do show an association with DTF-LD, even after taking into account the effect of all other SNPs and phenotypes. Therefore, the {\tt netphenogeno()} function reduces the number of candidate SNPs and gives a small set of much more plausible ones. 
Moreover, \cite{balasubramanian2009qtl} have reported that the TLN-SD phenotype is associated with a region in chromosome 5, whereas our proposed method do not find any direct effect between TLN-SD and the region in chromosome 5, only through the DTF-SD phenotype.
Furthermore, associations between phenotype CLN-LD and markers $snp49$ and $snp50$ have remained undetected in the previous studies of this population.  This example was run in about 4 minutes on an Intel i7 laptop with 16 GB RAM. 

In short, unlike traditional QTL analysis, the proposed method goes beyond the bivariate testing of individual SNPs, which only look at marginal association, instead it uses a multivariate approach which includes all the SNPs and phenotypes simultaneously. 


\paragraph{Genotype-phenotype networks in maize.} The high-dimensional genotypic and phenotypic maize data used in this paper were downloaded from \url{www.panzea.org}. The data comprised three datasets: a genotype data, and two phenotype datasets from the flowering time \citep{buckler2009genetic} and the leaf architecture \citep{tian2011genome}. The SNP data included $1106$ genetic markers for $194$ diverse maize recombinant inbred lines, which were derived from a cross between B73 and B97 from the maize Nested Association Mapping (NAM) populations. The 194 maize lines were scored for their flowering time using days to silking (DS), days to anthesis (DA), and the anthesis-silking interval (ASI) phenotypes. The leaf related traits such as upper leaf angle (ULA), leaf length (LL) and leaf width (LW) were also measured for all 194 maize lines.

\begin{figure}[!t]
	\centering
	\includegraphics[width=0.7\textwidth]{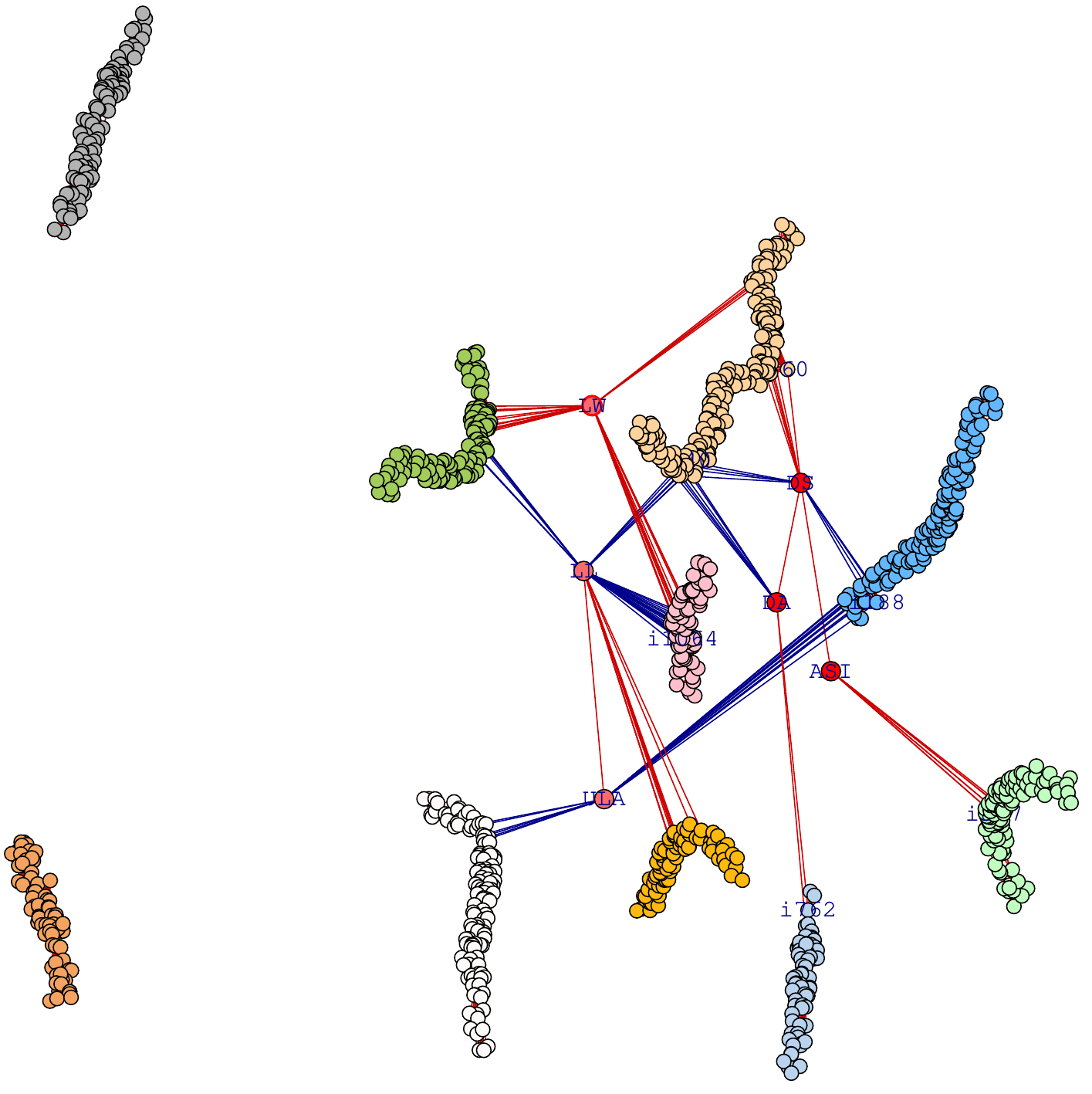}
	\caption{Genotype--phenotype networks for 1106 SNP markers and 6 phenotypes in mazie NAM population, where flowering traits (DS, DA, ASI) are shown in \tikzcircle[black, fill=red]{2.5pt} and leaf traits (LW, LL, ULA) are in \tikzcircle[black, fill=chestnut]{2.5pt}, respectively. SNPs are shown on chromosome 1 (snp1 - snp175) as \tikzcircle[black, fill=apricot]{2.5pt}, chromosome 2 (snp176 - snp302) as \tikzcircle[black, fill=iceberg]{2.5pt}, chromosome 3 (snp303 - snp432) as \tikzcircle[black, fill=applegreen]{2.5pt}, chromosome 4 (snp433 - snp543) as \tikzcircle[black, fill=gray]{2.5pt}, chromosome 5 (snp544 - snp682) as \tikzcircle[black, fill=white]{2.5pt}, chromosome 6 (snp683 - snp760) as \tikzcircle[black, fill=bronze]{2.5pt}, chromosome 7 (snp761 - snp838) as \tikzcircle[black, fill=beaublue]{2.5pt}, chromosome 8 (snp839 - snp944) as \tikzcircle[black, fill=celadon]{2.5pt}, chromosome 9 (snp945 - snp1029) as \tikzcircle[black, fill=amber]{2.5pt}, and chromosome 10 (snp1030 - snp1106) as \tikzcircle[black, fill=bubblegum]{2.5pt}. Blue edges show negative and red positive partial correlations. 
	} 
	\label{maize}
\end{figure}

Fig \ref{maize} reconstructs genotype--phenotype networks between the 6 phenotypes and 1106 SNPs. Five SNPs on chromosome 1 (from i140 until i144) directly affect both DA and DS traits (related to flowering time) after removing the effect of other variables.   
Moreover, a few SNPs on chromosome 1 (from i60 until i64) and on the beginning of chromosome 2 (i188 until i191) regulate DS.
Two SNP markers (i762 and i763) on chromosome 7 affect DA, and chromosome 8 (i877 until i883) regulates ASI phenotype, after adjustments.
The two leaf related traits, ULA and LL, are linked together, but not to the LW. Three SNPs i1064, i1062, and i1080 are yet conditionally associated to both LL and LW traits after adjustments.
Chromosomes 4 and 6 do not have any role in the studied flowering time and leaf architecture traits.

\begin{figure}[t]
	\centering
	\includegraphics[width=0.7\textwidth]{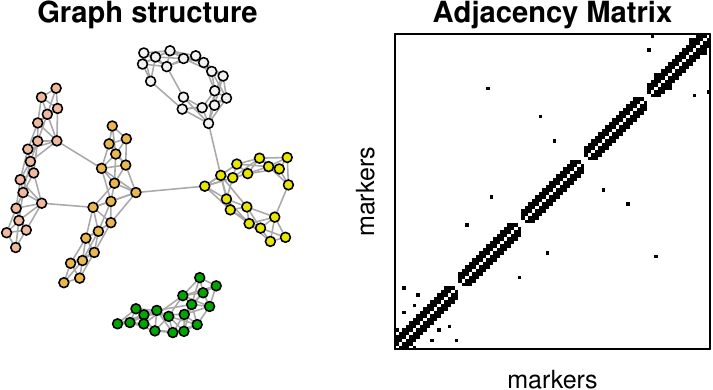}
	\caption{An example simulated (genotype) data using the function {\tt simgeno()}. (left) Each node presents an SNP marker and the colors correspond to a given number of linkage groups, (right) the correspondent adjacency matrix. The connectivity between the pairs of non-adjacent markers in a same linkage group can be controlled via the argument {\tt alpha} and the inter-chromosomal edges with {\tt beta}.	
	} 
\label{FigSim}   
\end{figure}

\subsection{Simulations and computational timing} 
The package generates simulated data in two ways
\begin{enumerate} 
	\item {\tt simgeno()} function simulates genotype data based on a Gaussian copula graphical model. An inbred genotype data can be generated for {\tt p} number of SNP markers, for {\tt n} number of individuals, for {\tt k} genotype states in a q-ploid species where $q$ represents chromosome copy number (or ploidy level of chromosomes). The simulated data mimic a genome-like graph structure: First, there are {\tt g} linkage groups (each of which represents a chromosome); then within each linkage group adjacent markers, {\tt adjacent}, are linked via an edge as a result of genetic linkage. Also, with probability {\tt alpha} a pair of non-adjacent markers in the same chromosome are given an edge. Inter-chromosomal edges are simulated with probability {\tt beta}. These links represent long-range linkage disequilibriums. The corresponding positive definite precision matrix $\Theta$ has a zero pattern corresponding to the non-present edges. The underlying variable vector $Z$ is simulated from either a multivariate normal distribution, $N_p (0, \Theta^{-1})$, or a multivariate t-distribution with degrees of freedom {\tt d} and covariance matrix $\Theta^{-1}$. We generate the genotype marginals using random cutoff-points from a uniform distribution, and partition the latent space into {\tt k} states. The function can be called with the following arguments

	\begin{verbatim}
		set.seed(2)
		sim <- simgeno(p = 90, n = 200, k = 3, g = 5, adjacent = 3, alpha = 0.1, 
		beta = 0.02, con.dist = "Mnorm", d = NULL, vis = TRUE)
	\end{verbatim}	
	
	The output of the example is shown in Figure \ref{FigSim}.
	\item {\tt simRIL()} function generates diploid recombinant inbred lines (RILs) using recombination fraction and a CentiMorgan position of markers across the chromosomes. The function can be called with the following arguments
	
	\begin{verbatim}
		set.seed(2)
		ril <- simRIL(g = 5, d = 25, n = 200, cM = 100, selfing = 2)
		ril$data[1:3, ]
       M1.1 M2.1 M3.1 M4.1 M5.1 M6.1 M7.1 M8.1  ... M24.5 M25.5 
		ind1    0    0    0    0    0    0    0    0    ...   0     0          
		ind2    2    1    1    1    1    1    2    2    ...   1     1         
		ind3    1    2    2    2    2    2    2    1    ...   0     0
	\end{verbatim}

	\begin{verbatim}
		ril$map
		     chr  marker  cM
		1     1   M1.1   0.000000
		2     1   M2.1   4.166667
		.
		. 
		.
		124   5   M24.5  95.833333
		125   5   M25.5  100.00000
	\end{verbatim}
	
	where {\tt g} and {\tt d} represent the number of chromosomes and the number of markers in each chromosome, respectively. The number of sample size can be specified by {\tt n}. The arguments {\tt cM} and {\tt selfing} show the length of chromosome based on centiMorgan position and the number of selfing in the RIL population, respectively. 
\end{enumerate}

\begin{figure}[h]
	\centering
	\includegraphics[width=0.45\textwidth]{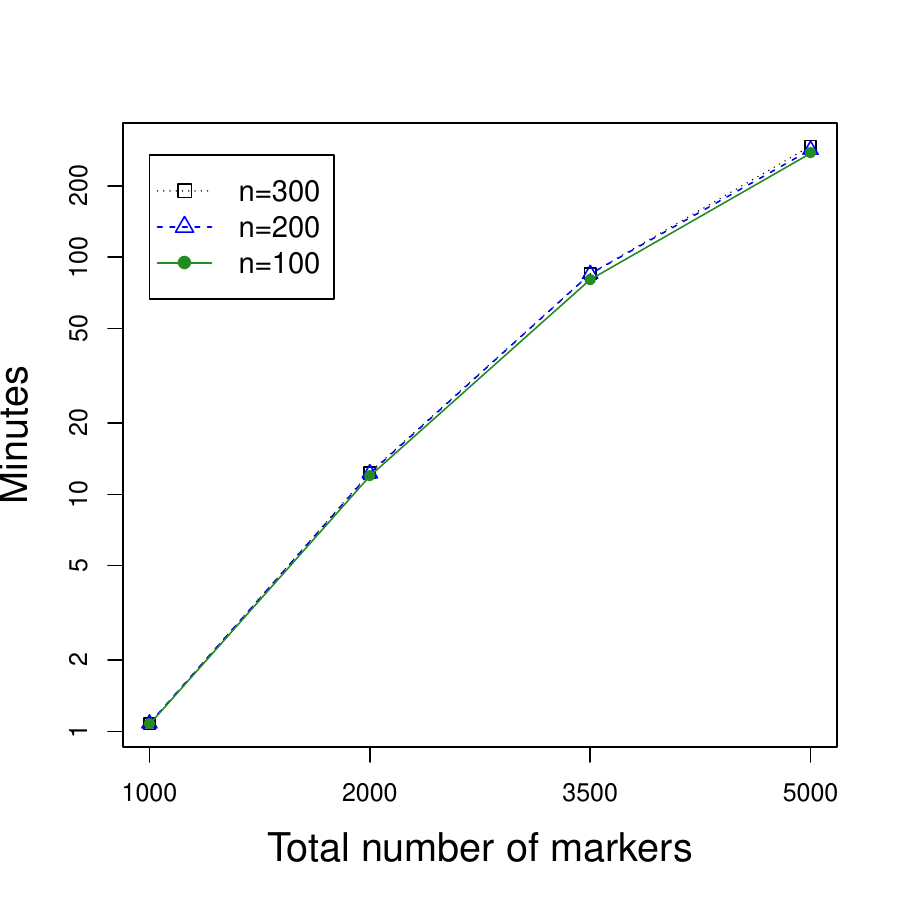} 
	\vspace{-0.4cm}
	\caption{The computational time of linkage map construction in \pkg{netgwas} for various simulated data with different combinations of individuals $n$ and the number of markers $p$, where they were distributed evenly across 10 linkage groups. This shows that the computational time is roughly proportional to $p^3$, as long as $p \times \max\{n,p\}$ elements can be stored in memory.} 
	\label{timingPlot}
\end{figure}

\paragraph{Computational timing.} Fig \ref{timingPlot} shows computational timing of \pkg{netgwas} for different number of variables $p$ and different sample sizes $n$. In this figure, we report computational timing in minutes for the genetic map construction,  which includes the graph estimation procedure and the ordering algorithm. Note that the other two functions  {\texttt netsnp()} and {\texttt nethenogeno()} include only the graph estimation, so we have only considered  {\texttt netmap()} function to cover the computational aspect of the \pkg{netgwas} package. For the simulated data, we generated $p=1000, 2000, 3500, 5000$ markers using {\texttt simRIL} function, which evenly are distributed across 10 chromosomes, for different individuals $n=100, 200, 300$.  
Fig \ref{timingPlot} shows that computational time is not affected by sample size n and is roughly proportional to $p^3$, as long as $p \times \max\{n,p\}$ elements can be stored in memory. The reported timing is based on the result from a computer with an Intel Core i7--6700 CPU and 32GB RAM. 

\section{Conclusion and future directions}
\label{Discussion}

The \pkg{netgwas} package implements the methods developed by \cite{behrouzi2019novo} and \cite{behrouzi2019detecting} to (i) construct linkage maps for bi-parental species with any ploidy level, namely diploid ($2$ sets), triploid ($3$ sets), tetraploid ($4$ sets) and so on; (ii) explore high--dimensional short-- and long--range linkage disequilibrium (LD) networks among pairs of SNP markers while controlling for the effect of other SNPs. The inferred LD networks reveal epistatic interactions across a genome when viability of the particular genetic recombination of the parental lines is considered as phenotype; 
(iii) infer genotype-phenotype networks from multi--loci multi--trait data, where it measures the pairwise associations with adjusting for the effect of other markers and phenotypes. Moreover, it detects markers that directly are responsible for that phenotype (disease), and reports the strength of their associations in terms of partial correlations. 
In addition, the package is able to reconstruct conditional dependence networks among SNPs, phenotypes, and environmental variables. 

The implemented method is based on copula graphical models that enables us to infer conditional independence networks from incomplete non-Gaussian data, ordinal data, and mixed ordinal-and-continuous data. The package uses a parallelization strategy on multi-core processors to speed-up computations for large datasets. In addition, the code is memory-optimized, using the sparse matrix data structure when estimating and storing full regularization paths for large data sets. The \pkg{netgwas} package contains several functions for simulation and interactive network visualization. We note that reproducibility of our results and all the example data used to illustrate the package is supported by the open-source R package \pkg{netgwas}.

The \pkg{netgwas} and \pkg{qtl2} \citep{broman2019r} software are for high-dimensional genotype and phenotype data. The \pkg{qtl2} performs QTL analysis in multi-parental populations solely by genome scans with single-QTL models. The function {\tt netphnogeno} in \pkg{netgwas} uses a multivariate approach to detect conditional interactions networks between genotypes and phenotypes. It uses network models to detect multiple causal SNPs in a QTL region in bi-parental populations, while adjusting for the effect of remaining QTLs. One of the primary directions for the future work is to extend our methodology for multi-parental map construction and perform their QTL analysis. This would require the calculation of genotype probabilities using hidden Markov models \citep{broman2009guide, zheng2018accurate} before implementing the proposed Gaussian copula graphical model. 

We will maintain and develop the package further. In the future we will include time component into our model where the interest is to infer dynamic networks for longitudinal (phenotyping) data and to learn changes of networks over time. Implementation of such model is desirable in many fields, particularly in plant breeding where the main goal is to optimize yield using high-throughput phenotypic data. 

\section*{Acknowledgment}
The authors are grateful to the associated editor and reviewers for their valuable comments that improved the manuscript and the R package.

\bibliography{netgwasRef}

\begin{thebibliography}{68}
\providecommand{\natexlab}[1]{#1}
\providecommand{\url}[1]{\texttt{#1}}
\expandafter\ifx\csname urlstyle\endcsname\relax
  \providecommand{\doi}[1]{doi: #1}\else
  \providecommand{\doi}{doi: \begingroup \urlstyle{rm}\Url}\fi

\bibitem[Balasubramanian et~al.(2009)Balasubramanian, Schwartz, Singh,
  Warthmann, Kim, Maloof, Loudet, Trainer, Dabi, Borevitz,
  et~al.]{balasubramanian2009qtl}
S.~Balasubramanian, C.~Schwartz, A.~Singh, N.~Warthmann, M.~C. Kim, J.~N.
  Maloof, O.~Loudet, G.~T. Trainer, T.~Dabi, J.~O. Borevitz, et~al.
\newblock Qtl mapping in new arabidopsis thaliana advanced
  intercross-recombinant inbred lines.
\newblock \emph{PLoS One}, 4\penalty0 (2):\penalty0 e4318, 2009.

\bibitem[Bates and Maechler(2014)]{Matrix}
D.~Bates and M.~Maechler.
\newblock \emph{Matrix: Sparse and Dense Matrix Classes and Methods}, 2014.
\newblock URL \url{http://CRAN.R-project.org/package=Matrix}.
\newblock R package version 1.1-2.

\bibitem[Behrouzi and Wit(2019{\natexlab{a}})]{behrouzi2019detecting}
P.~Behrouzi and E.~C. Wit.
\newblock Detecting epistatic selection with partially observed genotype data
  by using copula graphical models.
\newblock \emph{Journal of the Royal Statistical Society: Series C (Applied
  Statistics)}, 68\penalty0 (1):\penalty0 141--160, 2019{\natexlab{a}}.

\bibitem[Behrouzi and Wit(2019{\natexlab{b}})]{behrouzi2019novo}
P.~Behrouzi and E.~C. Wit.
\newblock De novo construction of polyploid linkage maps using discrete
  graphical models.
\newblock \emph{Bioinformatics}, 35\penalty0 (7):\penalty0 1083--1093,
  2019{\natexlab{b}}.

\bibitem[Behrouzi et~al.(2018)Behrouzi, Abegaz, and Wit]{behrouzi2018dynamic}
P.~Behrouzi, F.~Abegaz, and E.~C. Wit.
\newblock Dynamic chain graph models for ordinal time series data.
\newblock \emph{ArXiv preprint ArXiv:1805.09840}, 2018.

\bibitem[Bikard et~al.(2009)Bikard, Patel, Le~Mette, Giorgi, Camilleri,
  Bennett, and Loudet]{bikard2009divergent}
D.~Bikard, D.~Patel, C.~Le~Mette, V.~Giorgi, C.~Camilleri, M.~J. Bennett, and
  O.~Loudet.
\newblock Divergent evolution of duplicate genes leads to genetic
  incompatibilities within a. thaliana.
\newblock \emph{Science}, 323\penalty0 (5914):\penalty0 623--626, 2009.

\bibitem[Bourke et~al.(2018)Bourke, van Geest, Voorrips, Jansen, Kranenburg,
  Shahin, Visser, Arens, Smulders, and Maliepaard]{bourke2018polymapr}
P.~M. Bourke, G.~van Geest, R.~E. Voorrips, J.~Jansen, T.~Kranenburg,
  A.~Shahin, R.~G. Visser, P.~Arens, M.~J. Smulders, and C.~Maliepaard.
\newblock polymapr—linkage analysis and genetic map construction from f1
  populations of outcrossing polyploids.
\newblock \emph{Bioinformatics}, 34\penalty0 (20):\penalty0 3496--3502, 2018.

\bibitem[Broman(2009)]{broman2009brief}
K.~W. Broman.
\newblock A brief tour of r/qtl.
\newblock \emph{Disponivel http://www. rqtl. org/tutorials/rqtltour. pdf},
  2009.

\bibitem[Broman and Sen(2009)]{broman2009guide}
K.~W. Broman and S.~Sen.
\newblock \emph{A Guide to QTL Mapping with R/qtl}, volume~46.
\newblock Springer, 2009.

\bibitem[Broman et~al.(2003)Broman, Wu, Sen, and Churchill]{broman2003r}
K.~W. Broman, H.~Wu, S.~Sen, and G.~A. Churchill.
\newblock R/qtl: Qtl mapping in experimental crosses.
\newblock \emph{Bioinformatics}, 19\penalty0 (7):\penalty0 889--890, 2003.

\bibitem[Broman et~al.(2019)Broman, Gatti, Simecek, Furlotte, Prins, Sen,
  Yandell, and Churchill]{broman2019r}
K.~W. Broman, D.~M. Gatti, P.~Simecek, N.~A. Furlotte, P.~Prins, {\'S}.~Sen,
  B.~S. Yandell, and G.~A. Churchill.
\newblock R/qtl2: software for mapping quantitative trait loci with
  high-dimensional data and multiparent populations.
\newblock \emph{Genetics}, 211\penalty0 (2):\penalty0 495--502, 2019.

\bibitem[Buckler et~al.(2009)Buckler, Holland, Bradbury, Acharya, Brown,
  Browne, Ersoz, Flint-Garcia, Garcia, Glaubitz, et~al.]{buckler2009genetic}
E.~S. Buckler, J.~B. Holland, P.~J. Bradbury, C.~B. Acharya, P.~J. Brown,
  C.~Browne, E.~Ersoz, S.~Flint-Garcia, A.~Garcia, J.~C. Glaubitz, et~al.
\newblock The genetic architecture of maize flowering time.
\newblock \emph{Science}, 325\penalty0 (5941):\penalty0 714--718, 2009.

\bibitem[Bush and Moore(2012)]{bush2012chapter}
W.~S. Bush and J.~H. Moore.
\newblock Chapter 11: Genome-wide association studies.
\newblock \emph{PLoS Computational Biology}, 8\penalty0 (12):\penalty0
  e1002822, 2012.

\bibitem[Butte et~al.(2000)Butte, Tamayo, Slonim, Golub, and
  Kohane]{butte2000discovering}
A.~J. Butte, P.~Tamayo, D.~Slonim, T.~R. Golub, and I.~S. Kohane.
\newblock Discovering functional relationships between rna expression and
  chemotherapeutic susceptibility using relevance networks.
\newblock \emph{Proceedings of the National Academy of Sciences}, 97\penalty0
  (22):\penalty0 12182--12186, 2000.

\bibitem[Buzdugan et~al.(2016)Buzdugan, Kalisch, Navarro, Schunk, Fehr, and
  B{\"u}hlmann]{buzdugan2016assessing}
L.~Buzdugan, M.~Kalisch, A.~Navarro, D.~Schunk, E.~Fehr, and P.~B{\"u}hlmann.
\newblock Assessing statistical significance in multivariable genome wide
  association analysis.
\newblock \emph{Bioinformatics}, 32\penalty0 (13):\penalty0 1990--2000, 2016.

\bibitem[Clarke et~al.(2011)Clarke, Anderson, Pettersson, Cardon, Morris, and
  Zondervan]{clarke2011basic}
G.~M. Clarke, C.~A. Anderson, F.~H. Pettersson, L.~R. Cardon, A.~P. Morris, and
  K.~T. Zondervan.
\newblock Basic statistical analysis in genetic case-control studies.
\newblock \emph{Nature Protocols}, 6\penalty0 (2):\penalty0 121--133, 2011.

\bibitem[Cuthill and McKee(1969)]{cuthill1969reducing}
E.~Cuthill and J.~McKee.
\newblock Reducing the bandwidth of sparse symmetric matrices.
\newblock In \emph{proceedings of the 1969 24th national conference}, pages
  157--172. ACM, 1969.

\bibitem[Dobra et~al.(2004)Dobra, Hans, Jones, Nevins, Yao, and
  West]{dobra2004sparse}
A.~Dobra, C.~Hans, B.~Jones, J.~R. Nevins, G.~Yao, and M.~West.
\newblock Sparse graphical models for exploring gene expression data.
\newblock \emph{Journal of Multivariate Analysis}, 90\penalty0 (1):\penalty0
  196--212, 2004.

\bibitem[Edwards et~al.(2010)Edwards, De~Abreu, and
  Labouriau]{edwards2010selecting}
D.~Edwards, G.~C. De~Abreu, and R.~Labouriau.
\newblock Selecting high-dimensional mixed graphical models using minimal aic
  or bic forests.
\newblock \emph{BMC Bioinformatics}, 11\penalty0 (1):\penalty0 1--13, 2010.

\bibitem[Foygel and Drton(2010)]{foygel2010extended}
R.~Foygel and M.~Drton.
\newblock Extended bayesian information criteria for gaussian graphical models.
\newblock \emph{Advances in neural information processing systems}, 23, 2010.

\bibitem[Friedman et~al.(2008)Friedman, Hastie, and
  Tibshirani]{friedman2008sparse}
J.~Friedman, T.~Hastie, and R.~Tibshirani.
\newblock Sparse inverse covariance estimation with the graphical lasso.
\newblock \emph{Biostatistics}, 9\penalty0 (3):\penalty0 432--441, 2008.

\bibitem[Friedman(2004)]{friedman2004inferring}
N.~Friedman.
\newblock Inferring cellular networks using probabilistic graphical models.
\newblock \emph{Science}, 303\penalty0 (5659):\penalty0 799--805, 2004.

\bibitem[Geman and Geman(1984)]{geman1984stochastic}
S.~Geman and D.~Geman.
\newblock Stochastic relaxation, gibbs distributions, and the bayesian
  restoration of images.
\newblock \emph{IEEE Transactions on Pattern Analysis and Machine
  Intelligence}, \penalty0 (6):\penalty0 721–741, 1984.

\bibitem[Grandke et~al.(2017)Grandke, Ranganathan, van Bers, de~Haan, and
  Metzler]{grandke2017pergola}
F.~Grandke, S.~Ranganathan, N.~van Bers, J.~R. de~Haan, and D.~Metzler.
\newblock Pergola: Fast and deterministic linkage mapping of polyploids.
\newblock \emph{BMC Bioinformatics}, 18\penalty0 (1):\penalty0 12, 2017.

\bibitem[Green(1990)]{green1990use}
P.~J. Green.
\newblock On use of the em for penalized likelihood estimation.
\newblock \emph{Journal of the Royal Statistical Society. Series B
  (Methodological)}, pages 443--452, 1990.

\bibitem[Hackett et~al.(2017)Hackett, Boskamp, Vogogias, Preedy, and
  Milne]{hackett2017tetraploidsnpmap}
C.~A. Hackett, B.~Boskamp, A.~Vogogias, K.~F. Preedy, and I.~Milne.
\newblock Tetraploidsnpmap: software for linkage analysis and qtl mapping in
  autotetraploid populations using snp dosage data.
\newblock \emph{Journal of Heredity}, 108\penalty0 (4):\penalty0 438--442,
  2017.

\bibitem[Hartemink et~al.(2000)Hartemink, Gifford, Jaakkola, and
  Young]{hartemink2000using}
A.~J. Hartemink, D.~K. Gifford, T.~S. Jaakkola, and R.~A. Young.
\newblock Using graphical models and genomic expression data to statistically
  validate models of genetic regulatory networks.
\newblock In \emph{Biocomputing 2001}, pages 422--433. World Scientific, 2000.

\bibitem[Hastings(1970)]{hastings1970monte}
W.~K. Hastings.
\newblock Monte carlo sampling methods using markov chains and their
  applications.
\newblock 1970.

\bibitem[He and Lin(2010)]{he2010variable}
Q.~He and D.-Y. Lin.
\newblock A variable selection method for genome-wide association studies.
\newblock \emph{Bioinformatics}, 27\penalty0 (1):\penalty0 1--8, 2010.

\bibitem[Hedrick(1987)]{hedrick1987gametic}
P.~W. Hedrick.
\newblock Gametic disequilibrium measures: proceed with caution.
\newblock \emph{Genetics}, 117\penalty0 (2):\penalty0 331--341, 1987.

\bibitem[Hoff(2007)]{hoff2007extending}
P.~D. Hoff.
\newblock Extending the rank likelihood for semiparametric copula estimation.
\newblock \emph{The Annals of Applied Statistics}, 1\penalty0 (1):\penalty0
  265--283, 2007.

\bibitem[Hoggart et~al.(2008)Hoggart, Whittaker, De~Iorio, and
  Balding]{hoggart2008simultaneous}
C.~J. Hoggart, J.~C. Whittaker, M.~De~Iorio, and D.~J. Balding.
\newblock Simultaneous analysis of all snps in genome-wide and re-sequencing
  association studies.
\newblock \emph{PLoS Genetics}, 4\penalty0 (7):\penalty0 e1000130, 2008.

\bibitem[Hsieh et~al.(2011)Hsieh, Dhillon, Ravikumar, and
  Sustik]{hsieh2011sparse}
C.-J. Hsieh, I.~S. Dhillon, P.~K. Ravikumar, and M.~A. Sustik.
\newblock Sparse inverse covariance matrix estimation using quadratic
  approximation.
\newblock In \emph{Advances in Neural Information Processing Systems}, pages
  2330--2338, 2011.

\bibitem[Huang et~al.(2012)Huang, Shah, George, et~al.]{huang2012dlmap}
B.~E. Huang, R.~Shah, A.~W. George, et~al.
\newblock dlmap: An r package for mixed model qtl and association analysis.
\newblock \emph{Journal of Statistical Software}, 50\penalty0 (6):\penalty0
  1--22, 2012.

\bibitem[Jordan(2004)]{jordan2004graphical}
M.~I. Jordan.
\newblock Graphical models.
\newblock \emph{Statistical Science}, 19\penalty0 (1):\penalty0 140--155, 2004.

\bibitem[Kaler et~al.(2020)Kaler, Gillman, Beissinger, and
  Purcell]{kaler2020comparing}
A.~S. Kaler, J.~D. Gillman, T.~Beissinger, and L.~C. Purcell.
\newblock Comparing different statistical models and multiple testing
  corrections for association mapping in soybean and maize.
\newblock \emph{Frontiers in Plant Science}, 10:\penalty0 1794, 2020.

\bibitem[Kang et~al.(2008)Kang, Zaitlen, Wade, Kirby, Heckerman, Daly, and
  Eskin]{kang2008efficient}
H.~M. Kang, N.~A. Zaitlen, C.~M. Wade, A.~Kirby, D.~Heckerman, M.~J. Daly, and
  E.~Eskin.
\newblock Efficient control of population structure in model organism
  association mapping.
\newblock \emph{Genetics}, 178\penalty0 (3):\penalty0 1709--1723, 2008.

\bibitem[Kang et~al.(2010)Kang, Sul, Service, Zaitlen, Kong, Freimer, Sabatti,
  and Eskin]{kang2010variance}
H.~M. Kang, J.~H. Sul, S.~K. Service, N.~A. Zaitlen, S.-y. Kong, N.~B. Freimer,
  C.~Sabatti, and E.~Eskin.
\newblock Variance component model to account for sample structure in
  genome-wide association studies.
\newblock \emph{Nature Genetics}, 42\penalty0 (4):\penalty0 348--354, 2010.

\bibitem[Klaassen and Wellner(1997)]{klaassen1997efficient}
C.~A. Klaassen and J.~A. Wellner.
\newblock Efficient estimation in the bivariate normal copula model: normal
  margins are least favourable.
\newblock \emph{Bernoulli}, pages 55--77, 1997.

\bibitem[Klasen et~al.(2016)Klasen, Barbez, Meier, Meinshausen, B{\"u}hlmann,
  Koornneef, Busch, and Schneeberger]{klasen2016multi}
J.~R. Klasen, E.~Barbez, L.~Meier, N.~Meinshausen, P.~B{\"u}hlmann,
  M.~Koornneef, W.~Busch, and K.~Schneeberger.
\newblock A multi-marker association method for genome-wide association studies
  without the need for population structure correction.
\newblock \emph{Nature Communications}, 7:\penalty0 13299, 2016.

\bibitem[Kruijer et~al.(2020)Kruijer, Behrouzi, Bustos-Korts,
  Rodr{\'\i}guez-{\'A}lvarez, Mahmoudi, Yandell, Wit, and van
  Eeuwijk]{kruijer2020reconstruction}
W.~Kruijer, P.~Behrouzi, D.~Bustos-Korts, M.~X. Rodr{\'\i}guez-{\'A}lvarez,
  S.~M. Mahmoudi, B.~Yandell, E.~Wit, and F.~A. van Eeuwijk.
\newblock Reconstruction of networks with direct and indirect genetic effects.
\newblock \emph{Genetics}, 214\penalty0 (4):\penalty0 781--807, 2020.

\bibitem[Lander et~al.(1987)Lander, Green, Abrahamson, Barlow, Daly, Lincoln,
  and Newburg]{lander1987mapmaker}
E.~S. Lander, P.~Green, J.~Abrahamson, A.~Barlow, M.~J. Daly, S.~E. Lincoln,
  and L.~Newburg.
\newblock Mapmaker: an interactive computer package for constructing primary
  genetic linkage maps of experimental and natural populations.
\newblock \emph{Genomics}, 1\penalty0 (2):\penalty0 174--181, 1987.

\bibitem[Lauritzen(1996)]{lauritzen1996graphical}
S.~Lauritzen.
\newblock \emph{Graphical Models}, volume~17.
\newblock Oxford University Press, USA, 1996.

\bibitem[Lauritzen and Sheehan(2003)]{lauritzen2003graphical}
S.~L. Lauritzen and N.~A. Sheehan.
\newblock Graphical models for genetic analyses.
\newblock \emph{Statistical Science}, pages 489--514, 2003.

\bibitem[Lincoln and Lander(1992)]{lincoln1992systematic}
S.~E. Lincoln and E.~S. Lander.
\newblock Systematic detection of errors in genetic linkage data.
\newblock \emph{Genomics}, 14\penalty0 (3):\penalty0 604--610, 1992.

\bibitem[Lippert et~al.(2011)Lippert, Listgarten, Liu, Kadie, Davidson, and
  Heckerman]{lippert2011fast}
C.~Lippert, J.~Listgarten, Y.~Liu, C.~M. Kadie, R.~I. Davidson, and
  D.~Heckerman.
\newblock Fast linear mixed models for genome-wide association studies.
\newblock \emph{Nature Methods}, 8\penalty0 (10):\penalty0 833--835, 2011.

\bibitem[Liu et~al.(2012)Liu, Han, Yuan, Lafferty, Wasserman,
  et~al.]{liu2012high}
H.~Liu, F.~Han, M.~Yuan, J.~Lafferty, L.~Wasserman, et~al.
\newblock High-dimensional semiparametric gaussian copula graphical models.
\newblock \emph{The Annals of Statistics}, 40\penalty0 (4):\penalty0
  2293--2326, 2012.

\bibitem[Mangin et~al.(2012)Mangin, Siberchicot, Nicolas, Doligez, This, and
  Cierco-Ayrolles]{mangin2012novel}
B.~Mangin, A.~Siberchicot, S.~Nicolas, A.~Doligez, P.~This, and
  C.~Cierco-Ayrolles.
\newblock Novel measures of linkage disequilibrium that correct the bias due to
  population structure and relatedness.
\newblock \emph{Heredity}, 108\penalty0 (3):\penalty0 285, 2012.

\bibitem[Margarido et~al.(2007)Margarido, Souza, and
  Garcia]{margarido2007onemap}
G.~Margarido, A.~Souza, and A.~Garcia.
\newblock Onemap: Software for genetic mapping in outcrossing species.
\newblock \emph{Hereditas}, 144\penalty0 (3):\penalty0 78--79, 2007.

\bibitem[Massa et~al.(2015)Massa, Manrique-Carpintero, Coombs, Zarka, Boone,
  Kirk, Hackett, Bryan, and Douches]{massa2015genetic}
A.~N. Massa, N.~C. Manrique-Carpintero, J.~J. Coombs, D.~G. Zarka, A.~E. Boone,
  W.~W. Kirk, C.~A. Hackett, G.~J. Bryan, and D.~S. Douches.
\newblock Genetic linkage mapping of economically important traits in
  cultivated tetraploid potato (solanum tuberosum l.).
\newblock \emph{G3: Genes, Genomes, Genetics}, 5\penalty0 (11):\penalty0
  2357--2364, 2015.

\bibitem[Metropolis et~al.(1953)Metropolis, Rosenbluth, Rosenbluth, Teller, and
  Teller]{metropolis1953equation}
N.~Metropolis, A.~W. Rosenbluth, M.~N. Rosenbluth, A.~H. Teller, and E.~Teller.
\newblock Equation of state calculations by fast computing machines.
\newblock \emph{The Journal of Chemical Physics}, 21\penalty0 (6):\penalty0
  1087–1092, 1953.

\bibitem[Newman(2004)]{newman2004fast}
M.~E. Newman.
\newblock Fast algorithm for detecting community structure in networks.
\newblock \emph{Physical review E}, 69\penalty0 (6):\penalty0 066133, 2004.

\bibitem[Panagiotou et~al.(2011)Panagiotou, Ioannidis, and
  Project]{panagiotou2011should}
O.~A. Panagiotou, J.~P. Ioannidis, and G.-W.~S. Project.
\newblock What should the genome-wide significance threshold be? empirical
  replication of borderline genetic associations.
\newblock \emph{International Journal of Epidemiology}, 41\penalty0
  (1):\penalty0 273--286, 2011.

\bibitem[Rakitsch et~al.(2012)Rakitsch, Lippert, Stegle, and
  Borgwardt]{rakitsch2012lasso}
B.~Rakitsch, C.~Lippert, O.~Stegle, and K.~Borgwardt.
\newblock A lasso multi-marker mixed model for association mapping with
  population structure correction.
\newblock \emph{Bioinformatics}, 29\penalty0 (2):\penalty0 206--214, 2012.

\bibitem[Sammon(1969)]{sammon1969nonlinear}
J.~W. Sammon.
\newblock A nonlinear mapping for data structure analysis.
\newblock \emph{IEEE Transactions on computers}, 100\penalty0 (5):\penalty0
  401--409, 1969.

\bibitem[Simon et~al.(2008)Simon, Loudet, Durand, Berard, Brunel, Sennesal,
  Durand-Tardif, Pelletier, and Camilleri]{simon2008qtl}
M.~Simon, O.~Loudet, S.~Durand, A.~Berard, D.~Brunel, F.~Sennesal,
  M.~Durand-Tardif, G.~Pelletier, and C.~Camilleri.
\newblock Qtl mapping in five new large ril populations of arabidopsis thaliana
  genotyped with consensus snp markers.
\newblock \emph{Genetics}, 178:\penalty0 2253--2264, 2008.

\bibitem[Taylor and Butler(2017)]{taylor2017r}
J.~Taylor and D.~Butler.
\newblock R package asmap: Efficient genetic linkage map construction and
  diagnosis.
\newblock \emph{Journal of Statistical Software}, 79\penalty0 (6), 2017.

\bibitem[Taylor et~al.(2011)Taylor, Verbyla, et~al.]{taylor2011r}
J.~Taylor, A.~Verbyla, et~al.
\newblock R package wgaim: Qtl analysis in bi-parental populations using linear
  mixed models.
\newblock \emph{Journal of Statistical Software}, 40\penalty0 (7):\penalty0
  1--18, 2011.

\bibitem[Tian et~al.(2011)Tian, Bradbury, Brown, Hung, Sun, Flint-Garcia,
  Rocheford, McMullen, Holland, and Buckler]{tian2011genome}
F.~Tian, P.~J. Bradbury, P.~J. Brown, H.~Hung, Q.~Sun, S.~Flint-Garcia, T.~R.
  Rocheford, M.~D. McMullen, J.~B. Holland, and E.~S. Buckler.
\newblock Genome-wide association study of leaf architecture in the maize
  nested association mapping population.
\newblock \emph{Nature Genetics}, 43\penalty0 (2):\penalty0 159, 2011.

\bibitem[Vinciotti et~al.(2022)Vinciotti, Behrouzi, and
  Mohammadi]{vinciotti2022bayesian}
V.~Vinciotti, P.~Behrouzi, and R.~Mohammadi.
\newblock Bayesian structural learning of microbiota systems from count
  metagenomic data.
\newblock \emph{arXiv preprint arXiv:2203.10118}, 2022.

\bibitem[Wang et~al.(2016)Wang, van Eeuwijk, and Jansen]{wang2016potential}
H.~Wang, F.~A. van Eeuwijk, and J.~Jansen.
\newblock The potential of probabilistic graphical models in linkage map
  construction.
\newblock \emph{Theoretical and Applied Genetics}, pages 1--12, 2016.

\bibitem[Welter et~al.(2013)Welter, MacArthur, Morales, Burdett, Hall, Junkins,
  Klemm, Flicek, Manolio, Hindorff, et~al.]{welter2013nhgri}
D.~Welter, J.~MacArthur, J.~Morales, T.~Burdett, P.~Hall, H.~Junkins, A.~Klemm,
  P.~Flicek, T.~Manolio, L.~Hindorff, et~al.
\newblock The nhgri gwas catalog, a curated resource of snp-trait associations.
\newblock \emph{Nucleic Acids Research}, 42\penalty0 (D1):\penalty0
  D1001--D1006, 2013.

\bibitem[Wu et~al.(2002)Wu, Ma, Painter, and Zeng]{wu2002simultaneous}
R.~Wu, C.-X. Ma, I.~Painter, and Z.-B. Zeng.
\newblock Simultaneous maximum likelihood estimation of linkage and linkage
  phases in outcrossing species.
\newblock \emph{Theoretical Population Biology}, 61\penalty0 (3):\penalty0
  349--363, 2002.

\bibitem[Wu et~al.(2008)Wu, Bhat, Close, and Lonardi]{wu2008efficient}
Y.~Wu, P.~R. Bhat, T.~J. Close, and S.~Lonardi.
\newblock Efficient and accurate construction of genetic linkage maps from the
  minimum spanning tree of a graph.
\newblock \emph{PLoS Genetics}, 4\penalty0 (10):\penalty0 e1000212, 2008.

\bibitem[Yang et~al.(2008)Yang, Hu, Hu, Yu, Xia, Ye, and
  Zhu]{yang2008qtlnetwork}
J.~Yang, C.~Hu, H.~Hu, R.~Yu, Z.~Xia, X.~Ye, and J.~Zhu.
\newblock Qtlnetwork: mapping and visualizing genetic architecture of complex
  traits in experimental populations.
\newblock \emph{Bioinformatics}, 24\penalty0 (5):\penalty0 721--723, 2008.

\bibitem[Yu et~al.(2006)Yu, Pressoir, Briggs, Bi, Yamasaki, Doebley, McMullen,
  Gaut, Nielsen, Holland, et~al.]{yu2006unified}
J.~Yu, G.~Pressoir, W.~H. Briggs, I.~V. Bi, M.~Yamasaki, J.~F. Doebley, M.~D.
  McMullen, B.~S. Gaut, D.~M. Nielsen, J.~B. Holland, et~al.
\newblock A unified mixed-model method for association mapping that accounts
  for multiple levels of relatedness.
\newblock \emph{Nature Genetics}, 38\penalty0 (2):\penalty0 203--208, 2006.

\bibitem[Zheng et~al.(2018)Zheng, Boer, and van Eeuwijk]{zheng2018accurate}
C.~Zheng, M.~P. Boer, and F.~A. van Eeuwijk.
\newblock Accurate genotype imputation in multiparental populations from
  low-coverage sequence.
\newblock \emph{Genetics}, 210\penalty0 (1):\penalty0 71--82, 2018.

\bibitem[Zych et~al.(2015)Zych, Li, van~der Velde, Joosen, Ligterink, Jansen,
  and Arends]{zych2015pheno2geno}
K.~Zych, Y.~Li, J.~K. van~der Velde, R.~V. Joosen, W.~Ligterink, R.~C. Jansen,
  and D.~Arends.
\newblock Pheno2geno-high-throughput generation of genetic markers and maps
  from molecular phenotypes for crosses between inbred strains.
\newblock \emph{BMC Bioinformatics}, 16\penalty0 (1):\penalty0 51, 2015.

\end{thebibliography}

\address{Pariya Behrouzi\\
  Applied Mathematics and Statistics - Biometris\\
  Wageningen University and Research\\
  6708PB, Wageningen\\
  The Netherlands\\
  (https://orcid.org/0000-0001-6762-5433)\\
  \email{pariya.behrouzi@wur.nl}}

\address{Danny Arends\\
  Albrecht Daniel Thaer-Institut für Agrar und Gartenbauwissenschaften\\
  Humboldt-Universität zu Berlin\\
  Invalidenstraße 42, 10115 Berlin\\
  Germany\\
  (https://orcid.org/0000-0001-8738-0162)\\
  \email{Danny.Arends@Gmail.com}}

\address{Ernst C. Wit\\
  Faculty of Science \& Informatics\\
  Universita della Svizzera Italiana (USI) \\
  Via Buffi 13\\
  6900 Lugano\\
  Switzerland\\
  (https://orcid.org/0000-0002-3671-9610)\\
  \email{wite@usi.ch}}

\end{article}

\end{document}